\documentstyle[aps,epsf]{revtex}

\begin{document}

\preprint{\parbox{6cm}{~ \hfil hep-ph/9903450\\
                       ~ \hfil IFT-P.006/99, January/99}
}

\title{GLUON FIELD STRENGTH CORRELATION FUNCTIONS WITHIN A CONSTRAINED INSTANTON
MODEL}
\author{A.~E.~Dorokhov$^{1,2}$, S.V. Esaibegyan$^{1,3}$, A.~E.~Maximov$^{1}$, and
S.V. Mikhailov$^{1}$}
\address{
$^{1}${\it Joint Institute for Nuclear Research,}\\
Bogoliubov Laboratory of Theoretical Physics,\\
141980, Moscow Region, Dubna, Russia\\
$^{2}${\it Instituto de Fisica Teorica, UNESP,}\\
Rua Pamplona, 145, 01405-900, Sao Paulo, Brazil\\
$^{3}${\it Yerevan Physics Institute, 375036, Yerevan, Armenia}
}
\date{\today }
\maketitle
\begin{abstract}
We suggest a constrained instanton (CI) solution in the physical QCD vacuum
which is described by large-scale vacuum field fluctuations. This solution
decays exponentially at large distances. It is stable only if the
interaction of the instanton with the background vacuum field is small and
additional constraints are introduced. The CI solution is explicitly
constructed in the ansatz form, and the two-point vacuum correlator of gluon
field strengths is calculated in the framework of the effective instanton
vacuum model. At small distances the results are qualitatively similar to
the single instanton case, in particular, the form factor $D_{1}$ is small,
which is in agreement with the lattice calculations. \newline
\end{abstract}

\vskip .1 cm

\section{Introduction}

\bigskip

The non-perturbative vacuum of QCD is densely populated by long - wave
fluctuations of gluon and quark fields. The order parameters of this
complicated state are characterized by the vacuum matrix elements of various
singlet combinations of quark and gluon fields, condensates: $\left\langle :%
\bar{q}q:\right\rangle $, ~$\left\langle :~F_{\mu \nu }^{a}F_{\mu \nu
}^{a}:\right\rangle $,$\left\langle :\bar{q}(\sigma _{\mu \nu }F_{\mu \nu
}^{a}\frac{\lambda ^{a}}{2})q:\right\rangle $ , {\it etc}. The nonzero quark
condensate $\left\langle :\bar{q}q:\right\rangle $ is responsible for the
spontaneous breakdown of chiral symmetry, and its value was estimated a long
time ago within the current algebra approach. The nonzero gluon condensate $%
\left\langle :~F_{\mu \nu }^{a}F_{\mu \nu }^{a}:\right\rangle $ through
trace anomaly provides the mass scale for hadrons, and its value was
estimated within the QCD sum rule (SR) approach. The importance of the QCD
vacuum properties for hadron phenomenology has been established by Shifman,
Vainshtein and Zakharov \cite{SVZ79}. They used the operator product
expansion to relate the behavior of hadron current correlation functions at
short distances to a small set of condensates. The values of low -
dimensional condensates were obtained phenomenologically from the QCD SR
analysis of various hadron channels.

Later on the nonlocal vacuum condensates or vacuum correlators have been
introduced \cite{NLC81,MihRad92}. They describe the distribution of quarks
and gluons in the non-perturbative vacuum. Physically, it means that vacuum
quarks and gluons can flow through the vacuum with nonzero momentum. From
this point of view the standard vacuum expectation values (VEVs) like $%
\left\langle :\bar{q}q:\right\rangle $, $\left\langle :\bar{q}%
D^{2}q:\right\rangle $~, $\left\langle :g^{2}F^{2}:\right\rangle ,\ \ldots $
appear as expansion coefficients of the quark $M(x)=\ \left\langle :~\bar{q}%
(0)\hat{E}(0,x)q(x):\right\rangle $ and gluon $D^{\mu \nu ,\rho \sigma }(x) $
correlators in a Taylor series in the variable $x^{2}/4$.

The correlator $D^{\mu \nu ,\rho \sigma }(x)$ of gluonic strengths\footnote{%
We follow the convention when the coupling constant is absorbed into the
gauge field $A_{\nu }(x)$.},
\begin{equation}
D^{\mu \nu ,\rho \sigma }(x-y)\equiv \left\langle :TrF^{\mu \nu }(x)\hat{E}%
(x,y)F^{\rho \sigma }(y)\hat{E}(y,x):\right\rangle ,  \label{GluCor}
\end{equation}
may be parameterized in the form consistent with general requirements of the
gauge and Lorentz symmetries as \cite{DoSi88}:
\begin{eqnarray}
D^{\mu \nu ,\rho \sigma }(x) &\equiv &\frac{1}{24}\left\langle
:F^{2}:\right\rangle \{(\delta _{\mu \rho }\delta _{\nu \sigma }-\delta
_{\mu \sigma }\delta _{\nu \rho })[D(x^{2})+D_{1}(x^{2})]+  \label{Fld_Cor}
\\
&+&(x_{\mu }x_{\rho }\delta _{\nu \sigma }-x_{\mu }x_{\sigma }\delta _{\nu
\rho }+x_{\nu }x_{\sigma }\delta _{\mu \rho }-x_{\nu }x_{\rho }\delta _{\mu
\sigma })\frac{\partial D_{1}(x^{2})}{\partial x^{2}}\},  \nonumber
\end{eqnarray}
where $\hat{E}(x,y)=P\exp \left( i\int_{x}^{y}A_{\mu }(z)dz^{\mu }\right) $
is the path-ordered Schwinger phase factor (the integration is performed
along the {\it straight} line) required for gauge invariance and $%
\displaystyle A_{\mu }(z)=A_{\mu }^{a}(z)\frac{\lambda ^{a}}{2}$, ~$%
\displaystyle F_{\mu \nu }(x)=F_{\mu \nu }^{a}(x)\frac{\lambda ^{a}}{2},$ $%
~F_{\mu \nu }^{a}(x)=\partial _{\mu }A_{\nu }^{a}(x)-\partial _{\nu }A_{\mu
}^{a}(x)+f^{abc}A_{\mu }^{b}(x)A_{\nu }^{c}(x)$. The $P-$exponential ensures
the parallel transport of color from one point to other. In (\ref{Fld_Cor}),
$\left\langle :F^{2}:\right\rangle =\left\langle :F_{\mu \nu }^{a}(0)F_{\mu
\nu }^{a}(0):\right\rangle $ is a gluon condensate, and $D(x^{2})$ and $%
D_{1}(x^{2})$ are form factors which characterize nonlocal properties of the
condensate in different directions. They are normalized at zero by the
conditions $D(0)=\kappa $, $D_{1}(0)=1-\kappa $, that depend on the dynamics
considered. For example, for the self-dual fields $\kappa =1$, while in the
Abelian theory without monopoles the Bianchi identity provides $\kappa =0$.

The gluon correlators $D_{\mu \nu \rho \sigma }(x)$ are involved in an
analysis \cite{Vol79} of the spectrum of bound states of heavy $Q\overline{Q}
$ systems. The level shift depends on the parameter $\lambda \tau $, where $%
\tau =4/m_{Q}\alpha _{s}^{2}$ is the typical time of the low lying levels of
the system, and $\lambda $ is the correlation length of the gluon correlator
$\lambda $ defined as $D(x\rightarrow \infty )\sim \left\langle
:F^{2}:\right\rangle \exp (-|x|/\lambda )$. Thus, at large distances the
physically motivated asymptotics of the correlator is exponentially
decreasing. The gluon correlators are the base elements of the stochastic
model of vacuum \cite{DoSi88} and in the description of high-energy hadron
scattering \cite{Nacht87}.

Measuring the correlation length and vacuum field correlators was the
motivation to investigate these quantities on the lattice. New
high-statistical LQCD measurements of the gauge - invariant bilocal
correlator of the gluon field strengths have become available down to a
distance of $0.1$ fm \cite{DiGi97}. Recently, the field strength correlators
have also been studied in the effective dual Abelian Higgs model in \cite
{BBDV} and QCD sum rule approach \cite{DEJ}. In all these approaches (see
also \cite{Bali97}), the exponential decay of the correlators at large
distances has been observed. However, these investigations still omit a
small and intermediate distances behavior of the nonlocal condensates.

On the other hand, in QCD there is an instanton \cite{BPST}, a well known
nontrivial nonlocal vacuum solution of the classical Euclidean Yang - Mills
equations with finite action and size $\rho $. The importance of instantons
for QCD is that it is believed that an interacting instanton ensemble
provides a realistic microscopic picture of the QCD vacuum in the form of
instanton liquid \cite{Sh82,DP84} (see, {\it e.g.}, a review \cite{Shuryak96}%
). It has been argued on phenomenological grounds that the distribution of
instantons over sizes is peaked at a typical value $\rho _{c}\approx 1.7\
GeV^{-1}$ and the liquid is dilute in the sense that mean separation between
instantons is much larger than the average instanton size.

In our previous work \cite{DEM97}, we have shown that the instanton model of
the QCD vacuum provides a way to construct nonlocal vacuum condensates.
Within the effective single instanton (SI) approximation we have obtained
the expressions for the nonlocal gluon $D_{I}^{\mu \nu ,\rho \sigma }(x)$
and quark $M_{I}(x)$ condensates and derived the average virtualities of
quarks $\lambda _{q}^{2}$ and gluons $\lambda _{g}^{2}$ in the QCD vacuum.
It has been found that due to specific properties of the SI approximation
(self-duality of the field) the $D_{1}$ gluon form factor is exactly zero.
The behavior of the correlation functions demonstrates that in the SI
approximation the model of nonlocal condensates can well reproduce the
behavior of the quark and gluon correlators at {\em short distances}.
Really, the quark and gluon average virtualities, defined via the first
derivatives of the nonlocal condensates $M_{I}(x^{2})$, ~$D_{I}(x)$ at the
origin,
\begin{equation}
\displaystyle\lambda _{q}^{2}\equiv -\frac{8}{M_{I}(0)}\frac{dM_{I}(x^{2})}{%
dx^{2}}\left| _{x=0}=2\frac{1}{\rho _{c}^{2}}\right. ,\qquad \lambda
_{g}^{2}\equiv -8\frac{dD_{I}(x^{2})}{dx^{2}}\left| _{x=0}=\frac{24}{5}\frac{%
1}{\rho _{c}^{2}}\right. ,  \label{Q_Virt}
\end{equation}
are connected with VEVs that parameterize the QCD SR,
\begin{equation}
\lambda _{q}^{2}\equiv \frac{\left\langle :\bar{q}D^{2}q:\right\rangle }{%
\left\langle :\bar{q}q:\right\rangle },\ \ \ \lambda _{g}^{2}\equiv \frac{%
\left\langle :F_{\mu \nu }^{a}\tilde{D}^{2}F_{\mu \nu }^{a}:\right\rangle }{%
\left\langle :F^{2}:\right\rangle }=2\frac{\left\langle
:fF^{3}:\right\rangle }{\left\langle :F^{2}:\right\rangle }-2\frac{%
\left\langle :g^{4}J^{2}:\right\rangle }{\left\langle :F^{2}:\right\rangle }.
\label{Q_Virt2}
\end{equation}
Where $\left\langle :fF^{3}:\right\rangle =\left\langle :f^{abc}F_{\mu \nu
}^{a}F_{\nu \rho }^{b}F_{\rho \mu }^{c}:\right\rangle $, $J^{2}=J_{\mu
}^{a}J_{\mu }^{a}$ and $J_{\mu }^{a}=\bar{q}(x)\frac{\lambda ^{a}}{2}\gamma
_{\mu }q(x)$. The value of $\lambda _{q}^{2}\approx 0.5\ {\rm \ GeV}^{2}$
estimated in the QCD SR analysis \cite{Piv91,BI82} is reproduced at $\rho
_{c}\approx 2\ {\rm GeV}^{-1}$. This number is close to the estimate from
the phenomenology of the QCD vacuum in the instanton liquid model. The model
provides parameterless prediction for the ratio $\lambda _{g}^{2}/\lambda
_{q}^{2}=12/5$ and so, $\lambda _{g}^{2}\approx 1.2\ {\rm \ GeV}^{2}$. In
\cite{DEM97} the effect of the inclusion of Schwinger exponent into
semiclassical calculations has been analyzed. For some quantities this
effect is very strong, being of an order of $100\%$ for gluon and quark
average virtualities.

Nevertheless, the SI approximation used evidently fails in the description
of physically argued distributions at large distances. In asymptotics, we
have found $M_{I}(x\rightarrow \infty )\sim \rho _{c}^{2}/x^{2}$ and $%
D_{I}(x\rightarrow \infty )\sim \rho _{c}^{4}/x^{4}$ for the quark and gluon
correlators, respectively. Thus, the SI solution over mathematical vacuum
provides too slow asymptotics at large distances.We should conclude that in
order to have a realistic model of vacuum correlators, the important effects
of instanton interaction with the long - wave vacuum configurations have to
be included.

The key point in the picture of realistic instanton vacuum is the
interaction of pseudoparticles in the vacuum. In \cite{CDG78}, the
interaction of a SI with an arbitrary weak external field has been examined
and dipole-dipole forces in a far separated instanton-anti-instanton system
derived. Later in \cite{SVZ80}, this background field has been interpreted
as a field of large-scale QCD vacuum fluctuations, and the influence of the
quark and gluon condensates on the instanton density has been considered. \
The main assumption of the instanton liquid models \cite{Sh82} is the
dominance of the instanton component in the vacuum and that, in
particularly, the gluon condensate is saturated by weakly interacting
instanton liquid. In deriving instanton ensemble properties the
instanton-anti-instanton interaction at intermediate separation start to
play the crucial role in stabilization of the liquid \cite{DP84}. However,
it turns out that the final result strongly depends on the field ansatz for
an instanton-anti-instanton configuration \cite{Verb91,Shuryak96}. Further,
in all instanton-anti-instanton ansatze suggested the influence of the
physical vacuum on an instanton profile function has not been taken into
account and the profile has only power decreasing asymptotics, which
contradicts the expectations concerning the vacuum field correlators.
Moreover, it is known that the instanton liquid is not responsible for
large-scale effects like confinement \cite{Shuryak96}. Another point is that
the instanton density $n_{c}$ in the instanton liquid models is normalized
by the value of the gluon condensate $\left\langle :\frac{\alpha _{s}}{\pi }%
~F_{\mu \nu }^{a}F_{\mu \nu }^{a}:\right\rangle =0.012$ {\rm GeV}$^{4}$
obtained in \cite{SVZ79} from an analysis of charmonium spectrum. More
recently in \cite{Naris95}, a detailed analysis based on charmonium,
bottonium and heavy-light mesons have led to a twice larger value of the
gluon condensate $\left\langle :\frac{\alpha _{s}}{\pi }~F_{\mu \nu
}^{a}F_{\mu \nu }^{a}:\right\rangle =0.023$ {\rm GeV}$^{4}$. Indefiniteness
in the normalization provides a window for existence of a large-scale field
component in the QCD vacuum along with short-scale instantons.

In the present work, assuming dominance of the weak interacting instanton
liquid in the QCD vacuum, we suggest that there is also a weak residual
component of the vacuum field with a large correlation length $R$ of the
order of the confinement size. We are going to show, assuming only very
general properties of a weak large-scale vacuum field, that it deforms an
instanton at large distances leading to exponentially decreasing asymptotics
of the instanton profile and the instanton induced vacuum field correlators.
The vacuum model considered is a two-phase one. The large-scale phase is
described by the background field and dominates at distances compared with
the confinement size. The short-scale phase is dominated by instantons and
is responsible for the spontaneous breaking of chiral symmetry and the
solution of the $U_{A}\left( 1\right) $ problem. The vacuum model suggested
reveals a definite non-locality mechanism in the framework of QCD. We shall
illustrate this by analyzing the vacuum gluon form factors $D(x)$ and $%
D_{1}(x)$. Unfortunately, the normalization of contributions from different
phases to the gluon condensate is not fixed by the instanton model and
remains as a free parameter, but the form of the correlation functions can
be described in detail. The latter is the main motivation of the present
work. The determination of the vacuum field correlators is one of the main
tasks of the theory in describing the non-perturbative dynamics at large
distances compatible with the typical hadron size.

The paper is organized as follows: in Section 2 the instanton field in the
background of weak large-scale vacuum fluctuations is considered; solutions
of the vacuum field equations for small and large distances are analyzed
separately and the constrained instanton solution interpolating two
asymptotics is suggested in an ansatz form. In Section 3, the space
coordinate behavior of the nonlocal correlator of the gluon field strengths
is found and the main asymptotics of the correlators $D(x)$, $D_{1}(x)$ at
large distances are derived. These asymptotics are driven by the strength
and correlation length of the large-scale vacuum fluctuations, rather than
by their form.

\section{Constrained Instantons in QCD Vacuum}

The classical Yang - Mills equations in the Euclidean space
\begin{equation}
D_{\mu }F_{\mu \nu }(x)=0,  \label{YM}
\end{equation}
where the covariant derivative is $D_{\mu }=\partial _{\mu }-iA_{\mu
}^{a}\tau _{a}/2$, have an instanton $(+)$ ($(-)$ is for an anti -
instanton) as a (anti) self-dual finite-action solution with topological
charge, $Q=\pm 1$:
\begin{equation}
A_{sing,\mu }^{a,\pm }(x;x_{0})=2O_{I}^{ab}\eta _{\mu \nu }^{b,\pm } \left (
x-x_{0}\right) _{\nu }\varphi _{g}^{I}(x-x_{0})\quad \quad \mbox{with}\ \ \
\ \varphi _{g}^{I}(x)=\frac{\rho ^{2}}{x^{2}(x^{2}+\rho ^{2})}~%
\mbox{(in
singular gauge)},  \label{Ainst}
\end{equation}
localized in size $\rho $. In (\ref{Ainst}), $x_{0}$ is the position of the
instanton center, $O_{I}^{ab}$ is the ortogonal matrix of instanton global
orientation in the color space and $\displaystyle\eta _{\mu \nu }^{a,\pm
}=\epsilon _{4a\mu \nu }\mp \frac{1}{2}\epsilon _{abc}\epsilon _{bc\mu \nu }$
are t'Hooft symbols. The solution (\ref{Ainst}) is written in the {\it %
singular} gauge within the $SU(2)$ subgroup (with generators $\frac{\tau _{a}%
}{2}$) of the $SU_{c}(3)$ theory. This classical field configuration
reflects the symmetries of the initial theory in terms of collective
variables corresponding to translational transformations, rotational
symmetry in color space and conform transformations.

The solution (\ref{Ainst}) is given in mathematical vacuum and has an
unpleasant property of a very slow decay at large distances noted in the
introduction. This situation is inadequate since the physical vacuum is not
empty but looks like a medium densely populated by large-scale vacuum field
fluctuations. In the background of the large scale fluctuations there are
developed non - perturbative fluctuations of a smaller size among those
instantons which play a dominating role. The long-wave gluon vacuum field,
which is the background for a selected instanton, may be of a more general
origin and phenomenologically can be parameterized by the vacuum correlation
functions of the gluon operators contributing to the corresponding nonlocal
condensates.

What is important is that at {\em random} (stochastic) background vacuum
field with {\em fixed} vacuum expectation values of singlet operators the
scale invariance of the effective theory is spoiled already at the
quasi-classical level and the instantons are {\em no longer exact solutions}
of the field equations and the Dirac operator has no zero modes.

The similar situation has been observed in the standard electroweak
Yang-Mills-Higgs model. There, the background of Higgs field with nonzero
vacuum value $\left\langle \varphi \right\rangle $ and coupling $\lambda $
affects the instanton configuration \cite{tH76}. In the presence of (even
small) effects violating scale invariance of the initial theory, instanton
solution does not exist at all. Nevertheless, as it was stated in \cite{tH76}
and fairly explained in \cite{Affl,Esp90,Wang94}, if the Higgs field is
rather weak and some additional constraints are introduced, there may be
constructed an approximate solution, so-called, constrained instanton (CI).
These constraints limit the degree of freedom along the size $\rho $
parameter. The constrained solution at small distances $|x|<<\sqrt{\lambda }%
\left\langle \varphi \right\rangle ^{-1}$ approximately has the form of an
instanton, and at large distances $|x|>>\sqrt{\lambda }\left\langle \varphi
\right\rangle ^{-1}>>\rho $ has the asymptotics of massive (gauge boson)
particle $exp(-g\left\langle \varphi \right\rangle |x|/\sqrt{\lambda })$. In
\cite{Affl}, it has also been noted that the gauge field propagator of the
CIs decays exponentially at large $\left| x\right| $ and thus does not
affect the long-range behavior of the theory.

The aim of this section is to show that an analogous phenomenon takes place
in QCD considering an instanton field $A_{\mu }^{I}(x)$ in the physical
vacuum. In distinction with the Yang-Mills-Higgs model, in pure QCD there is
no Higgs field from the beginning and it is the long-scale vacuum gluon
field, $b_{\mu }(x)$, that models a source perturbing an instanton at large
distances. The deep reason of this effect is in an existence of the quantum
anomaly in the trace of energy-momentum tensor \cite{MigAgCho85}. It will be
shown that the presence of this background field, characterized by its
vacuum expectation value $\left\langle \left( F_{b,\mu \nu }^{a}\right)
^{2}\right\rangle _{b}$ and correlation length $R$ (introduced below), sets
the final scale and defines the deformation of the instanton solution in the
asymptotic region $|x|\gtrsim \left[ \left\langle \left( F_{b,\mu \nu
}^{a}\right) ^{2}\right\rangle _{b}R\right] ^{-1/3}>>\rho $. Here and below,
$\left\langle ...\right\rangle _{b}=\int d\sigma \left[ b\right] ...$ means
the average over nonperturbative random background field weighted with some
measure $d\sigma \left[ b\right] $. This solution is stable against
shrinking the instanton to a point if some constraints are added. By analogy
with the solutions analyzed in the Yang-Mills-Higgs model \cite{Affl}, we
shell call these interpolating fields constrained (or deformed) instanton
solutions.

In the following we analyze the vacuum field configuration of the single
(constrained) instanton $A_{\mu }^{CI}(x)$ of fixed size and orientation in
the color space in the background of the large-scale topologically neutral
random vacuum field $b_{\mu }(x)$\footnote{%
The similar model has been considered earlier in \cite{Agas86}.}
\begin{equation}
A_{\mu }(x)=A_{\mu }^{CI}(x,x_{0})+b_{\mu }(x),  \label{Atot}
\end{equation}
with gauge transformation property
\begin{equation}
A_{\mu }(x)\rightarrow U^{\dagger }\left( x\right) A_{\mu }(x)U\left(
x\right) +iU^{\dagger }\left( x\right) \partial _{\mu }U\left( x\right) ,
\label{GaugeTr1}
\end{equation}
where $U\left( x\right) $ is a gauge transformation matrix. Considering the
field ansatz (\ref{Atot}), one has to take the instanton field $A_{\mu
}^{CI}(x)$ in the {\it singular }gauge \cite{DP84,Shuryak96}
\begin{eqnarray}
&&A_{sing,\mu }^{CI,a,\pm }(x)=2O_{I}^{ab}\eta _{\mu \nu }^{b,\pm }\ \left(
x-x_{0}\right) _{\nu }\varphi _{g}^{CI}(x-x_{0}),  \label{A_CI} \\
&&\left. x^{2}\varphi _{g}^{CI}(x)\right| _{x\rightarrow 0}=1,\qquad \left.
\varphi _{g}^{CI}(x)\right| _{x\rightarrow \infty }=0.  \nonumber
\end{eqnarray}
The last conditions mean that the constrained instanton has a finite action
and a modulo unit topological charge, but, in general, ceases to be
self-dual field. In the coordinate space the instantons in this specific
gauge fall off rapidly enough to provide a weak interaction with the
background field and the quasi-classical approach is justified. The weakness
of the interaction allow us in the following to neglect the back reaction of
the instanton on the background field. As to nonperturbative background
field it is convenient to choose it in the Fock-Schwinger gauge \cite{D-Sm}
\begin{equation}
\left( x-x_{0}\right) _{\mu }b^{\mu }\left( x-x_{0}\right) =0.
\label{F-Schb}
\end{equation}
In the following we put the instanton center  $x_{0}$ to the origin of
coordinates $x_{0}=0$.

As an illustrative model for the background field, one can keep in mind the
self-dual homogeneous vacuum gluon field $b_{\mu }^{a}\left( x\right) =\frac{%
1}{2}n^{a}b_{\mu \nu }x_{\nu },\quad b_{\mu \nu }b_{\mu \nu }=b^{2}$, where $%
n^{a}$ is the orientation vector in color space and $b_{\mu \nu }$ is the
constant field-strength tensor \cite{Leut81}. The corresponding measure $%
\int d\sigma \left[ b\right] =\int_{0}^{\infty }dbD\left( b\right) \int
d\Omega \int d\Omega _{c}$ averages over field amplitude and its
orientations in configuration and color spaces. This field configuration
with infinite correlation length $R=\infty $ and infinite topological charge
quite correctly describes situation at small and intermediate distances
comparable with instanton size, but at larger distances the effect of finite
correlation length of physical background becomes important. Introduction of
finite correlation length can be imagined as the inclusion of domain
structure in the vacuum \cite{NachR84}. This kind of considerations are in
the base of the stochastic vacuum model \cite{DoSi88}. In the absence of a
consistent theoretical approach to the large distance dynamics one is led to
elaborate the problem phenomenologically.

We assume that at small distances the CI field dominates and the background
field $b_{\mu }\left( x\right) $ is regarded as a perturbation on $A_{\mu
}^{CI}(x)$. At distances much larger compared to the instanton size $\rho $
the background field $b_{\mu }\left( x\right) $ is still weak, but strong
enough to deform and suppress the instanton field.

The field strength can be written as
\begin{equation}
F_{\mu \nu }^{a}[A^{CI}+b]=F_{\mu \nu }^{CI,a}+F_{b,\mu \nu }^{a}+\Delta
F_{\mu \nu }^{a}\left[ A^{CI},b\right] ,  \label{Ftot}
\end{equation}
where $F_{\mu \nu }^{CI,a}\equiv F_{\mu \nu }^{a}[A^{CI}]$, $F_{b,\mu \nu
}^{a}\equiv F_{\mu \nu }^{a}[b]$ and
\[
\Delta F_{\mu \nu }^{a}\left[ A^{CI},b\right] =f_{abc}(A_{\mu }^{CI,b}b_{\nu
}^{c}+A_{\nu }^{CI,c}b_{\mu }^{b}),
\]
and the effective Euclidean action of the instanton in the {\em random}
background field becomes

\begin{equation}
S_{E}\approx \frac{1}{4g^{2}}\left\langle \int d^{4}x\left\{ F_{\mu \nu
}^{CI,a}F_{\mu \nu }^{CI,a}+\Delta F_{\mu \nu }^{a}\left[ A^{CI},b\right]
\Delta F_{\mu \nu }^{a}\left[ A^{CI},b\right] \right\} \right\rangle _{b}.
\label{SE}
\end{equation}
In deriving these expression we have used the color-singlet properties of
the large-scale vacuum on an average
\begin{equation}
\left\langle F_{b,\mu \nu }^{a}\right\rangle _{b}=0\   \label{ColAv}
\end{equation}
and average over relative orientation of instanton and background field in
color space. We also neglected the terms higher order in interaction.
Below, these terms will effectively be accumulated in the form of constraints below
and, what is important in the present consideration, they do not influence
the form of the solution at asymptotics.

Similar to Affleck analysis \cite{Affl}, we come to the conclusion that for
a background field configuration $b_{\mu }(x)$ with a {\em fixed} nonzero
condensate value no instanton solution exists. This can be seen from the
rescaling $x\rightarrow ax$, $A_{\mu }^{CI}(x)$ $\rightarrow a^{-1}A_{\mu
}^{CI}(ax)$, $b_{\mu }(x )$ $\rightarrow b_{\mu }(ax)$, preserving finite
vacuum average $\left\langle F_{b,\mu \nu }^{2}\right\rangle _{b}=const$,
under which $S_{E}$ transforms to
\begin{equation}
S_{E}\rightarrow \frac{1}{4g^{2}}\left\langle \int d^{4}x\left\{ F_{\mu \nu
}^{CI,a}F_{\mu \nu }^{CI,a}+a^{-2}\Delta F_{\mu \nu }^{a}\left[ A^{CI},b%
\right] \Delta F_{\mu \nu }^{a}\left[ A^{CI},b\right] \right\} \right\rangle
_{b}  \label{SEscl}
\end{equation}

If $A_{\mu }^{CI}(x)$ is a stationary point, then $dS_{E}/da=0$ and the
action is minimized by the instanton of vanishing size. Thus, given any
field configuration we can always rescale it to get smaller action, except
in the trivial case.

Now, let us consider the problem from the point of view of the equations of
motion for the deformed instanton in the background of large-scale random
vacuum fluctuations, that follows from ~(\ref{SE}),
\begin{equation}
D_{\mu }^{ab}\left[ A^{CI}\right] F_{\mu \nu }^{CI,b}+f^{bac}f^{bkl}\left(
A_{\mu }^{CI,k}\left\langle b_{\mu }^{c}b_{\nu }^{l}\right\rangle
_{b}-A_{\nu }^{CI,k}\left\langle b_{\mu }^{c}b_{\mu }^{l}\right\rangle
_{b}\right) =0.  \label{E-LeqAv}
\end{equation}
In the Fock-Schwinger gauge the background field has a representation in
terms of its strength
\begin{equation}
b_{\mu }^{a}(x)=\int_{0}^{1}d\alpha \alpha F_{b,\nu \mu }^{a}(\alpha
x)x^{\nu }  \label{F-Sch}
\end{equation}
and the bilinear field averages become
\begin{equation}
\left\langle b_{\mu }^{a}(x)b_{\nu }^{b}(x)\right\rangle
_{b}=\int_{0}^{1}d\alpha \int_{0}^{1}d\beta \alpha \beta x_{\rho }x_{\sigma
}\left\langle F_{b,\rho \mu }^{a}(\alpha x)F_{b,\sigma \nu }^{b}(\beta
x)\right\rangle _{b}.  \label{F-SchBB}
\end{equation}
In the non-Abelian case the correlator in the integrand of (\ref{F-SchBB})
is not gauge-invariant, however, in the Fock-Schwinger gauge this correlator
coincides with the gauge-invariant correlator in which field-strengths are
connected by the Schwinger phase factors $\tilde{E}(\alpha x,0)\tilde{E}%
(0,\beta x)$ in the adjoint representation. Thus in this specific gauge the
gauge variant left side of (\ref{F-SchBB}) can be expressed in terms of
gauge invariant quantity $\left\langle F_{b,\rho \mu }^{a}(\alpha x)\tilde{E}%
(\alpha x,\beta x))F_{b,\sigma \nu }^{b}(\beta x)\right\rangle _{b}$. Due to
the gauge-invariance the latter correlator admits a physically motivated
model
\begin{equation}
x_{\rho }x_{\sigma }\left\langle F_{\rho \mu }^{a}(\alpha x)F_{\sigma \nu
}^{b}(\beta x)\right\rangle _{b}=\frac{\delta ^{ab}}{N_{c}^{2}-1}\frac{%
\left\langle F_{b}^{2}\right\rangle _{b}}{12}\left( x^{2}\delta _{\mu \nu
}-x_{\mu }x_{\nu }\right) \left. \widetilde{B}\left( z^{2}\right) \right|
_{z=x\left( \alpha -\beta \right) },  \label{BB_Lor}
\end{equation}
where the function
\begin{equation}
\widetilde{B}\left( z^{2}\right) =\widetilde{D}\left( z^{2}\right) +%
\widetilde{D}_{1}\left( z^{2}\right) +z^{2}\partial \widetilde{D}_{1}\left(
z^{2}\right) /\partial z^{2}  \label{CorrLB}
\end{equation}
is defined via the form factors $\widetilde{D}\left( z^{2}\right) $ and $%
\widetilde{D_{1}}\left( z^{2}\right) $ parameterizing the gauge-invariant
two-point correlator (\ref{Fld_Cor}) of the background field strengths, with
normalization $\widetilde{D}(0)=\widetilde{\kappa }$, $\widetilde{D_{1}}%
\left( 0\right) =1-\widetilde{\kappa }$. The contribution of the background
field to the gluon condensate is denoted by $\left\langle
F_{b}^{2}\right\rangle _{b}$. With these definitions the equations of motion
of the CI field interacting with a random large-scale vacuum fluctuation
field, (\ref{E-LeqAv}), can be cast in the form
\begin{equation}
D_{\mu }^{ab}\left[ A^{CI}\right] F_{\mu \nu }^{CI,b}\left( x\right) -\frac{%
N_{c}\left\langle F_{b}^{2}\right\rangle _{b}}{24(N_{c}^{2}-1)}x^{2}\Phi
\left( x^{2}\right) A_{\mu }^{CI,a}\left( x\right) =0,  \label{E-LeqAv1}
\end{equation}
where
\begin{equation}
\Phi \left( x^{2}\right) =4\int_{0}^{1}d\alpha \int_{0}^{1}d\beta \alpha
\beta \widetilde{B}\left[ \left( \alpha -\beta \right) ^{2}x^{2}\right]
,\qquad \Phi \left( 0\right) =1,  \label{Phi}
\end{equation}
and $N_{c}$ is the number of colors.

Let us discuss the properties of the solution of Eq. (\ref{E-LeqAv1}). In
the absence of the background field $\left\langle F_{b}^{2}\right\rangle
_{b}=0$ there exists an instanton solution (\ref{Ainst}). For $\left\langle
F_{b}^{2}\right\rangle _{b}$ small enough, such that $\left\langle
F_{b}^{2}\right\rangle _{b}\ll 1/\rho ^{4}$, we should expect to find a
solution of (\ref{E-LeqAv1}) in perturbation theory in small parameter $%
\left\langle F_{b}^{2}\right\rangle _{b}\rho ^{4}$, which reduces to (\ref
{Ainst}) when $\left\langle F_{b}^{2}\right\rangle _{b}\rightarrow 0$.
However, such a perturbative solution does not exist. The reason is that for
the higher order in perturbation terms appropriate finite action boundary
conditions at large distances cannot be enforced \cite{Affl}.

The operators that act on higher order terms possess zero mode $\partial
A_{\mu }^{CI}/\partial \rho $
\begin{equation}
\nabla _{\mu }\left( \nabla _{\mu }\frac{\partial A_{\nu }^{CI}}{\partial
\rho }-\nabla _{\nu }\frac{\partial A_{\mu }^{CI}}{\partial \rho }\right) +i%
\left[ \frac{\partial A_{\mu }^{CI}}{\partial \rho },F_{\mu \nu }\left[
A^{CI}\right] \right] =0  \label{Z-Meq}
\end{equation}
which determines {\it a priori }the behavior of perturbative around
instanton terms at infinity. A way of getting around this difficulty \cite
{Affl} is to extremize the action $S_{E}$, (\ref{SE}), subject to a
constraint. The choice of an explicit form of the constraint is quite
arbitrary. In \cite{Affl} it has been proposed the global constraint of the
general form
\[
C_{constr}^{nl}\left( A\right) =\int d^{4}x\left[ O_{d}\left( A\right)
-O_{d}\left( A^{CI}\right) \right] =0,
\]
where the gauge-invariant local operator $O_{d}(A)$ has a canonical
dimension $d>4$. The relevant stationary configuration will be a solution of
the equations of motion (\ref{E-LeqAv1}) but with the constraint term added
into the right-hand side
\begin{equation}
\left. \frac{\delta S_{E}\left( A\right) }{\delta A_{\mu }\left( x\right) }%
\right| _{A_{\mu }^{CI}}=\sigma \left. \frac{\delta C_{constr}\left(
A\right) }{\delta A_{\mu }\left( x\right) }\right| _{A_{\mu }^{CI}}.
\label{Const_W}
\end{equation}

The Lagrange multiplier $\sigma $ in (\ref{Const_W}) is to be determined
order by order in perturbative theory in $\left\langle
F_{b}^{2}\right\rangle _{b}\rho ^{4}$, which provides the correct boundary
conditions for the higher order terms. The constrained instanton $A^{CI}$ is
the unique solution of (\ref{Const_W}) obtained by this procedure, the $%
A^{CI}$-solution turns to (\ref{Ainst}) when $\left\langle
F_{b}^{2}\right\rangle _{b}\rho ^{4}$ $\rightarrow 0$. Unfortunately, this
kind of constraints is nonlinear and the higher order terms in $\left\langle
F_{b}^{2}\right\rangle _{b}\rho ^{4}$, depending on the constraint, are
difficult to evaluate in practice.

Another way has been suggested in \cite{Wang94}, where a linear constraint
of the general form
\begin{equation}
C_{constr}^{l}\left( A\right) =\int d^{4}xtr\left\{ \left( A_{\mu }\left(
x\right) -A_{\mu }^{CI}\left( x\right) \right) f_{\mu }^{\rho }\left(
x\right) \right\} =0
\end{equation}
has been introduced. It has been proposed, \cite{Wang94}, instead of fixing
the constraint to solve the equation for $A_{\mu }^{CI}\left( x\right) $,
which is almost an impossible task, to choose $A_{\mu }^{CI}\left( x\right) $
first and then find the constraint $f_{\mu }^{\rho }\left( x\right) $ itself
by substituting $A_{\mu }^{CI}\left( x\right) $ into the left hand side of
the constrained equation (\ref{Const_W}). In this way, the freedom in
choosing the constraint can be used to find it by a given solution.

One can also restrict by considering only the local operators defining
constraints that fall down at infinity more rapidly than the interaction
term in Eq. (\ref{E-LeqAv1}). Under this condition, it is easy to obtain the
behavior of the instanton enveloped in the background field at distances far
from the instanton center. This large-distance asymptotics of CI, like its
behavior at small distances $A_{\mu }^{I}\left( x;\rho \right) $, is a
constraint-independent part of the solution.

We are interested in the asymptotic behavior of the function $\Phi \left(
x^{2}\right) $, Eq. (\ref{Phi}), where the interaction term becomes dominant
over the instanton self-interaction. It is nice that the leading asymptotics
of $\Phi \left( x^{2}\right)$, $\displaystyle \Phi \left( x^{2}\right) \sim
R/|x|$, where $R$ is a correlation length, is independent of a particular
form of the function $\widetilde{B} (z^2)$. This property is due to a
specific dependence of the argument $z^2= (\alpha-\beta)^2 x^2$. Let us
illustrate this property using a few natural ansatze for this function
\begin{eqnarray}  \label{B}
\widetilde{B}_{M}\left( x^{2}\right) =R^{2}/\left( x^{2}+R^{2}\right),& ~~%
\widetilde{B}_{E}\left( x^{2}\right) =\exp (-\left| x\right| /R),& ~~~%
\widetilde{B}_{G}\left( x^{2}\right) =\exp (-x^{2}/R^{2}), \\
\mbox{(Monopole)}~~~~~& \mbox{(Exponential)}& ~~~~\mbox {(Gaussian)}
\nonumber
\end{eqnarray}
that provide, respectively, (see appendix B)
\begin{equation}
\displaystyle\Phi^{as} \left( x^{2}\right) \stackrel{\left| x\right|
\rightarrow \infty }{\rightarrow }\frac{8}{3}a_{\Phi }\frac{R}{\left|
x\right| },\qquad \mbox{where}\qquad a_{\Phi }=O(1)=\left\{
\begin{array}{c}
\frac{\pi }{2}\qquad \mbox{monopole,} \\
1\qquad \mbox{exponential,} \\
\frac{\sqrt{\pi }}{2}\qquad \mbox{gaussian.}\qquad
\end{array}
\right.  \label{PhiAs}
\end{equation}
The exponential ansatz (modulo powers) appears as an asymptotics of the
solution of the Yang-Mills-Higgs model \cite{Affl,Esp90} and also is used in
parameterization of large-scale behavior of Lattice QCD data \cite{DiGi97}.
The monopole form resembles the asymptotic behavior of the SI correlator
\cite{DEM97}, where, in fact, it has steeper decrease like $1/x^{4}$.

Asymptotic behavior of the instanton solution deformed by large-scale vacuum
fluctuations at large Euclidean distances $|x|\rightarrow \infty $ can be
derived from the analysis of the equation
\begin{eqnarray}
&&\partial _{\mu }(\partial _{\mu }A_{\nu }^{CI}-\partial _{\nu }A_{\mu
}^{CI})-\eta _{g}^{3}\left| x\right| A_{\mu }^{CI}=0,  \label{E-LeqAvAs} \\
&&\eta _{g}=\left( \frac{a_{\Phi }N_{c}}{9\left( N_{c}^{2}-1\right) }
R\left\langle F_{b}^{2}\right\rangle _{b}\right) ^{\frac{1}{3}},
\end{eqnarray}
which follows from Eq.~\ref{E-LeqAv1}. Due to a decreasing character of the
field asymptotics, $A_{\mu }^{CI}\rightarrow 0$ at $|x|\rightarrow \infty $,
only linear terms in the short-wave CI field $A_{\mu }^{CI}$ are kept in the
kinetic part of equation (\ref{E-LeqAvAs})\footnote{%
At this point we have to note that the influence of the instanton ensemble
on the instanton profile has been discussed in \cite{DP84}. The authors
found that this interaction perturbs the self-interaction part by the term $%
\mu _{g}^{2}A_{\mu }$. It turns out that numerically the coefficient $\mu
_{g}^{2}$ strongly depends on the instanton-anti-instanton ansatz chosen
\cite{Verb91} and, as it is seen from (\ref{E-LeqAvAs}), has subleading
behavior in the limit of large $x$.}. For the profile function $\varphi
_{g}^{as}(x^{2})$ defined by ($\overline{\eta }_{\nu \mu }^{a}\equiv \eta
_{\nu \mu }^{+a}$ in the following)
\begin{equation}
A_{\mu }^{CI,a}(x)=\overline{\eta }_{\nu \mu }^{a}\frac{x_{\nu }}{x^{2}}%
\varphi _{g}^{as}\left( x^{2}\right) ,  \label{CIanz}
\end{equation}
we find from the asymptotic equation (\ref{E-LeqAvAs}) the large distance
solution
\begin{equation}
\varphi _{g}^{as}\left( x^{2}\right) \sim K_{4/3}\left[ \frac{2}{3}\left(
\eta _{g}\left| x\right| \right) ^{3/2}\right] ,  \label{AsSol}
\end{equation}
where $K_{\nu }(z)$ is the modified Bessel function with index $\nu $\
having the asymptotic behavior $K_{\nu }(z)\rightarrow \sqrt{\frac{\pi }{2z}}%
e^{-z}~~\mbox{as}~z\rightarrow \infty $. We have to note that in the case of
the homogeneous background field with infinite correlation length one get
the equation similar to (\ref{E-LeqAvAs}), but with the coefficient
proportional to $x^{2}$ in the last term that results in the Gaussian form
of asymptotics ({\it c.f.}, (\cite{MigAgCho85})).

The requirement of the instanton-background interaction  being weak  allows
us to consider the dimensionless parameter $\alpha _{g}\equiv \eta _{g}\rho $
a small value. It means that the region, where the instanton field dominates
(small distances), and the region, where the background field dominates
(large distances), are well separated and the large distance effects do not
destroy the instanton. Then, the overall constant is determined by matching,
at distances $\rho \ll \left| x\right| \ll \eta _{g}^{-1}$, the leading
terms of the expansions of $A_{\mu }^{CI}(x)$ at small distances, which is
an instanton $A_{\mu }^{I}(x)$, and at large distances, which is an
asymptotics (\ref{CIanz}), (\ref{AsSol}),
\begin{equation}
\displaystyle A_{\mu }^{CI,a}(x)=\overline{\eta }_{\nu \mu }^{a}\frac{%
2x_{\nu }}{x^{2}}a_{4/3}\alpha _{g}^{2}K_{4/3}\left[ \frac{2}{3}\left( \eta
_{g}\left| x\right| \right) ^{3/2}\right] ,\qquad \mbox{where}\qquad a_{4/3}=%
\frac{2}{\Gamma \left( 1/3\right) 3^{1/3}}  \label{CIas}
\end{equation}
is the normalization coefficient and $\Gamma (z)$ is the Gamma-function.

Thus, we find that in {\em the singular gauge} the CI solution has an
exponential decreasing character (\ref{CIas}) far from the instanton center.
This behavior sharply differs from the power decreasing asymptotics of SI (%
\ref{Ainst}). This modification of behavior follows from the fact that the
instanton solution is considered in the physical vacuum populated by the
large-scale gluon field fluctuations. The background field modifies the
long-distance behavior of the instanton and leads to appearance of the
``second scale'' parameter $\Lambda _{g}=1/\eta _{g}$ (see \cite{DEM97}) in
gluon distributions. At the same time, the effect of long-wave vacuum
fluctuations is not very essential for the behavior of the instanton at
short distances.

Now we are looking for the constrained solution of (\ref{E-LeqAv1}) in the
ansatz form
\begin{equation}
A_{\mu }^{CIa}(x)=2\overline{\eta }_{\nu \mu }^{a}x_{\nu }\varphi _{g}\left(
x^{2}\right)  \label{CIanA}
\end{equation}
which has the behavior at small and large distances
\begin{equation}
\displaystyle\varphi _{g}\left( x^{2}\right) =\frac{1}{x^{2}}\cdot \left\{
\begin{array}{c}
\displaystyle \frac{\rho ^{2}}{(x^{2}+\rho ^{2})}+\left( \eta _{g}\left|
x\right| \right) ^{3}\varphi _{\left( 1\right) }\left( \frac{\rho }{\left|
x\right| }\right) ...\ \ \ \ \ \mbox{at }\qquad |x|\rightarrow 0, \\
\displaystyle a_{4/3}\alpha _{g}^{2} K_{4/3}\left[ \frac{2}{3}\left( \eta
_{g}\left| x\right| \right) ^{3/2}\right] +\left( \frac{\rho ^{2}} {x^{2}}%
\right) ^{2}\varphi ^{\left( 1\right) }\left( \eta _{g}\left| x\right|
\right) ...\ \ \ \mbox{at }\quad |x|\rightarrow \infty ,
\end{array}
\right.  \label{CIas1}
\end{equation}
where the forms of the expansions can be deduced from expanding the leading
terms with respect to $\frac{\rho }{\left| x\right| }$ and $\eta _{g}\left|
x\right| $, respectively. The first terms in the expansions are exact and
constraint-independent ones, however, as it was shown in \cite{Affl}, all
higher order terms have dependence on the constraint chosen.

The condition of finiteness of the action, which is also
constraint-independent one, should be imposed on the desirable solution. To
this goal, let us rewrite the CI field strength as
\begin{equation}
F_{\mu \nu }^{CIa}\left( x\right) =4\left[ \overline{\eta }_{\mu \nu
}^{a}\omega _{1}\left( x\right) +\left( x_{\mu }\overline{\eta }_{\nu \rho
}^{a}-x_{\nu }\overline{\eta }_{\mu \rho }^{a}\right) x_{\rho }\omega
_{2}\left( x\right) \right]  \label{CIanF}
\end{equation}
with forms
\begin{equation}
\omega _{1}\left( x\right) =x^{2}\varphi _{g}^{2}\left( x^{2}\right)
-\varphi _{g}\left( x^{2}\right) ,\qquad \omega _{2}\left( x\right) =\varphi
_{g}^{2}\left( x^{2}\right) +\frac{\partial \varphi _{g}\left( x^{2}\right)
}{\partial x^{2}}, ~~~~~x_{\rho }^{2}=x^{2}+\rho ^{2}.  \label{OmFrms}
\end{equation}
to be used further. With this parameterization we have the action
\begin{equation}
S_{E}^{CI}=\frac{1}{4g^{2}}\int d^{4}x\left[ F_{\mu \nu }^{a}\left( x\right)
F_{\mu \nu }^{a}\left( x\right) \right] ^{CI}  \label{SCI}
\end{equation}
with the action density
\begin{equation}
\left [ F_{\mu \nu }^{a}\left( x\right) F_{\mu \nu }^{a}\left( x\right) %
\right]] ^{CI}=96\left[ \omega _{1}^{2}\left( x\right) +\omega
_{3}^{2}\left( x\right) \right] ,  \label{ActDens}
\end{equation}
where
\[
\omega _{3}\left( x\right) =x^{2}\omega _{2}\left( x\right) -\omega
_{1}\left( x\right) .
\]
Now, if we use the singular gauge for $A_{\mu }^{CI}$ , then, to guarantee
finiteness of the action, the condition has to be fulfilled
\begin{equation}
x^{2}\varphi _{g}\left( x^{2}\right) \left|_{x^{2}\rightarrow 0} \right.
\rightarrow 1+O\left( x^{2}\right) .\qquad \qquad  \label{FinAcCond}
\end{equation}
For further references we present here the well-known expressions for the SI
profiles in the singular and regular gauges
\begin{eqnarray}
\varphi _{g}^{sing,I}\left( x^{2}\right)& =& \frac{\rho^{2}}{x^{2}x_{\rho
}^{2} }, \qquad \qquad \varphi_{g}^{reg,I}\left( x^{2}\right) =\frac{1}{
x_{\rho }^{2}},  \label{InstSinReg} \\
\omega _{1}^{sing,I}\left( x\right) &=&-\frac{\rho ^{2}}{\left( x_{\rho
}^{2}\right) ^{2}},\qquad \qquad x^{2}\omega _{2}^{sing,I}\left( x\right)
=2\omega _{1}^{sing,I}\left( x\right) ,  \label{OmFrmsSing} \\
\omega _{1}^{reg,I}\left( x\right) &=&-\frac{\rho ^{2}}{\left( x_{\rho
}^{2}\right) ^{2}},\qquad \qquad \omega _{2}^{reg,I}\left( x\right) =0,
\label{OmFrmsReg}
\end{eqnarray}
and gauge-independent expressions for the density action and the action
itself
\begin{equation}
\left[ F_{\mu \nu }^{a}\left( x\right) F_{\mu \nu }^{a}\left( x\right) %
\right] ^{I}=\frac{192\rho ^{4}}{\left( x_{\rho }^{2}\right) ^{4}} ,\qquad
\qquad S_{E}^{I}=\frac{8\pi }{g^{2}}.  \label{InstAct}
\end{equation}
By using the asymptotic properties of the CI solution (\ref{CIas1}) and
finite-action condition (\ref{FinAcCond}), which are constraint-independent,
we are able to construct ansatz. Certainly, this procedure is not unique and
in principle one can impose further physical requirements to constrain the
behavior of the solution in the intermediate region. These details, however,
can be taken into account by choosing proper constraints. Thus, the freedom
in choosing the constraint can be used to find it by a given solution,
instead of solving complicated equations \cite{Wang94}.

Let us consider the following ansatze for the CI profile written in the
singular gauge
\begin{eqnarray}
\varphi _{1}\left( x^{2}\right) &=&\frac{1}{x^{2}}\frac{K_{4/3}\left[
z_{\rho ,x}\right] }{K_{4/3}\left[ z_{\rho ,0}\right] },  \label{CIprofM} \\
\varphi _{2}\left( x^{2}\right) &=&\frac{\overline{\rho }^{2}\left(
x^{2}\right) }{x^{2}x_{\rho }^{2}},  \label{CIprofW} \\
\varphi _{3}\left( x^{2}\right) &=&\frac{\overline{\rho }^{2}\left(
x^{2}\right) }{x^{2}\left( x^{2}+\overline{\rho }^{2}\left( x^{2}\right)
\right) },  \label{CIprofSW}
\end{eqnarray}
where we have introduced the notation
\begin{eqnarray*}
z_{\rho ,x} &=&\frac{2}{3}\eta _{g}^{3/2}\left( x^{2}+\rho ^{2}\right)
^{3/4}, \\
\overline{\rho }^{2}\left( x^{2}\right) &=&a_{4/3}\alpha
_{g}^{2}x^{2}K_{4/3} \left[ z_{0,x}\right] ,\qquad \overline{\rho }%
^{2}\left( 0\right) =\rho ^{2}.
\end{eqnarray*}
Note that all ansatze have easily identifiable instanton parameters. By
translational invariance the center of CI can be shifted in (\ref{CIprofM})
- (\ref{CIprofSW}) from the origin to an arbitrary position $x_{0} $: $%
x\rightarrow x-x_{0}$. The CI profile functions corresponding to these
ansatze are shown in Figs. (\ref{Fig1}) and (\ref{Fig2}) along with the
instanton profile (\ref{InstSinReg}). Figs. (\ref{Fig1}) and (\ref{Fig2})
display the dependence of the profile function on the external field
parameter $\rho \eta _{g}$ and on the form of ansatz, respectively. To make
the difference more clear, we also take for illustration a large value of
the parameter $\left( \rho \eta _{g}\right) ^{2}=3$. We see that if at small
distances the CI is close to the instanton form, then at large distances
this solution has exponential asymptotics instead of power-like for the
instanton.

Last two ansatze are similar to ones suggested in \cite{Wang94}, where the
preference has been given to the $\varphi _{3 }\left( x^{2}\right) $ form,
since it has better convergent properties in the expansion of CI (\ref{CIas1}%
). Moreover, with this profile the constrained solution in the regular gauge
looks similar to the SI case
\[
\varphi _{3}^{reg,CI}\left( x^{2}\right) =\frac{1}{x^{2}+\overline{\rho }%
^{2}\left( x^{2}\right) }.
\]
In order to pass from the constraint instanton in a singular gauge to the
instanton in a regular gauge one can translate the general gauge
transformation (\ref{GaugeTr1}) into the form
\begin{equation}
A_{\mu }^{CI}(x)\rightarrow \Omega ^{\dagger }\left( x\right) A_{\mu
}^{CI}(x)\Omega \left( x\right) +i\Omega ^{\dagger }\left( x\right) \partial
_{\mu }\Omega \left( x\right) ,  \label{Sin-RegTr}
\end{equation}
\[
b_{\mu }(x)\rightarrow \Omega ^{\dagger }\left( x\right) b_{\mu }(x)\Omega
\left( x\right) ,
\]
with transformation matrix
\[
\Omega \left( x\right) =\frac{i\tau _{\mu }^{-}x_{\mu }}{\left| x\right| }.
\]

We have numerically calculated the dependence of the CI classical action,
Eqs. (\ref{SCI}) and (\ref{ActDens}), on the instanton size $\rho $. This
dependence for the three ansatze (\ref{CIprofM}), (\ref{CIprofW}) and (\ref
{CIprofSW}) is shown in Figs. (\ref{Fig3}) and (\ref{Fig4}). We have to
stress, that the profiles of the field $A_{\mu }^{CI}$ and the action $%
S_{E}^{CI}$ depend on the choice of constraint. However, the full effective
action, with the terms coming from the Jacobian included, is
constraint-independent \cite{Wang94}. The weak dependence of the action, $%
S_{E}^{CI}$, on the profiles $\varphi _{1,2,3}\left( x^{2}\right) $ (see
Fig. (\ref{Fig4})) indicates that in the region of parameters $\rho \eta
_{g}\lesssim 1$, the influence of these additional terms is small and the
exponential part of the action, $S_{E}^{CI}$, can be used as a good
approximation. We see in Fig. (\ref{Fig4}) that the CI action is larger than
the instanton one and monotonically grows with the instanton size. It is
natural because the CI-``solution'' does not self-dual one and does not
realize the minimum of the action. Instead, it represents the bottom of the
valley parameterized by the quasi-zero mode $\rho $.

We are not going to discuss further details in construction of the total
effective CI action, that takes into account small quantum oscillations
around the nonperturbative configuration (\ref{Atot}) and the interaction of
constrained instantons, and postpone them until further publication. Just
point out that other effects dominating the effective action at small $\rho $
come from the running coupling constant $g^{2}\left( \rho \right) $ and the
path integral measure over the size of instanton $d\rho /\rho ^{5}$. It is
well known that in the model of the coupling constant, which freezes it to a
constant at some large $\rho _{0}$, the corrected action
\[
S_{E}^{TOT}\left( \rho \right) =\frac{1}{4g^{2}\left( \rho \right) }\int
d^{4}x\left[ F_{\mu \nu }^{a}\left( x\right) F_{\mu \nu }^{a}\left( x\right) %
\right] ^{CI}+5\ln \rho ,
\]
as a function of the instanton size, has a minimum. The position of the
minimum is correlated with the freezing parameter $\rho _{0}$, which can be
chosen to provide the value $\rho _{min}\approx 2$ {\rm GeV}$^{-1}$, and the
environment of the large-scale vacuum fluctuations makes the minimum more
prominent (see \cite{KogKov98} for recent discussion of similar results).
However, the typical CI action at $\rho _{min}$ is rather large, its
numerical value being around 25. This means that the configurations with
small number of instantons and anti-instantons are not important
statistically and suggests that the interacting instanton and anti-instanton
ensemble could be a more important type of configurations. The leading
interaction term of a widely separated instanton - anti-instanton pair in
physical vacuum as described by Eqs. (\ref{CIprofM}) - (\ref{CIprofSW})
falls with separation $L$ like $\displaystyle\exp \left( -4/3(\eta
_{g}L)^{3/2}\right) $ and differs from power-like decreasing behavior found
in \cite{CDG78,DP84} in unconstrained case. In the following, considering
the nonlocal properties of gluon condensate, we accept that the instanton
liquid is formed due to the instanton-anti-instanton interaction and the
instanton density $n_{c}$ and size $\rho _{c}$ are fixed \cite{Shuryak96}: $%
n_{c}\approx 1\quad {\rm fm}^{-4}$, $\rho _{c}\approx 1/3\quad {\rm fm}$.

\section{Short-range vacuum correlators in the constrained instanton model.}

\bigskip

Within the model considered the full gluon correlator may be conventionally
written by using the expression for the field strength (\ref{Ftot}) as
\begin{eqnarray}
&&\left\langle :F_{\mu \nu }[A^{CI}+b]\left( x\right) F_{\rho \sigma
}[A^{CI}+b]\left( y\right) :\right\rangle =\left\langle F_{\mu \nu
}^{CI}\left( x\right) F_{\rho \sigma }^{CI}\left( y\right) \right\rangle +
\nonumber \\
&&+\left\langle :F_{b,\mu \nu }\left( x\right) F_{b,\rho \sigma }\left(
y\right) :\right\rangle +\left\langle :\Delta F_{\mu \nu }[A^{CI},b]\left(
x\right) \Delta F_{\rho \sigma }[A^{CI},b]\left( y\right) :\right\rangle
^{interf},  \label{CorrFull}
\end{eqnarray}
where the brackets $\left\langle \quad \right\rangle $ mean averaging over
vacuum fluctuations (\ref{Atot}) and we do not display Schwinger phase
factors explicitly. The last term represents the interference of short- and
large-scale fields and will be discussed below. For the large-scale
correlator (\ref{CorrLB}) we have already suggested the model for the form
factor $\widetilde{B}(x^{2})$ in (\ref{B}). Now, let us calculate the
short-range part of the gluon correlator.

Let us construct the correlator $D^{\mu \nu ,\rho \sigma }(x-y)$ of gluonic
strengths (\ref{GluCor}) in the quasi-classical approximation by using the
CI solutions given by Eqs. (\ref{CIanA}) and (\ref{CIprofM}) - (\ref
{CIprofSW}). We will use a reference frame where the instanton sits at the
origin and a relative coordinate $\left( x-y\right) ^{\mu }$ with respect to
the position of the instanton center is parallel to one of the coordinate
axes, say $\mu =4$, serving as a ``time'' direction $({\it i.e.}, \vec{x}-
\stackrel{\rightarrow }{y}=0,\ x_{4}-y_{4}=|x-y|),$ and reduce the path
ordered exponential to an ordinary exponential
\begin{equation}
\hat{E}(x,y)=P\exp \left( i\int_{x}^{y}A_{\mu }^{CI}(z)dz^{\mu }\right)
=L^{\dagger }\left( x\right) L\left( y\right)  \label{SchwI}
\end{equation}
with
\begin{equation}
L\left( x\right) =\exp \left( \mp i\stackrel{\rightarrow }{\tau } \frac{%
\stackrel{\rightarrow }{x}}{\left| \stackrel{\rightarrow }{x}\right| }\alpha
\left( \left| \stackrel{\rightarrow }{x}\right| ,x_{4}\right) \right) = \mp
i\tau _{\mu }^{\pm }\cdot \widetilde{x}^{\mu }\left( x\right) ,  \label{R(x)}
\end{equation}
where
\begin{eqnarray}
&& \alpha \left( \left| \stackrel{\rightarrow }{x}\right| ,x_{4}\right)
=\left| \stackrel{\rightarrow }{x}\right| \int_{0}^{x_{4}}dt\varphi
_{g}\left( \left| \stackrel{\rightarrow }{x}\right| ^{2}+t^{2}\right) ,
\label{alphCI} \\
&&\tau^{\pm}= (\mp i,\vec{\tau}), ~\widetilde{x}^{0}\left( x\right) = \cos
\alpha \left( x\right) , ~~~\displaystyle\widetilde{x}^{i}\left( x\right)
=\left( x^{i}/\left| \stackrel{\rightarrow }{x}\right| \right) \sin \alpha
\left( x\right).
\end{eqnarray}
The factor $L\left( x\right) $ coming from the Schwinger exponent can be
accumulated in the definition of the field. This representation of the field
may be called the {\it axial }gauge representation $A_{\mu }(z)n^{\mu }=0,$
since in this gauge with the vector $n_{\mu }=x_{\mu }-y_{\mu }$ the
Schwinger factor $\hat{E}(x,y)=1$.

In the CI background the bilocal gluon correlator acquires the form
\begin{eqnarray}
D^{\mu \nu ,\rho \sigma }(x) &=&<:Tr\left( F_{(ax)\mu \nu }(0)F_{(ax)\rho
\sigma }(x)\right) :>=  \nonumber \\
\displaystyle &=&\sum_{\pm }n_{c}^{\pm }\int d^{4}{z}\int d\Omega Tr\left(
F_{(ax)\mu \nu }^{\pm }(z-\frac{x}{2})F_{(ax)\rho \sigma }^{\pm }(z+ \frac{x%
}{2})\right) ,  \label{Dgax}
\end{eqnarray}
where $n_{c}^{\pm }$ is the effective instanton / anti - instanton density, $%
z$ is the collective coordinate of the instanton center and $\Omega $ is its
color space orientation.

To extract form factors $D\left( x^{2}\right) $ and $D_{1}\left(
x^{2}\right) $ it is easier first to average over the instanton orientations
in the color space and take the trace over color matrices by using the
relations
\begin{eqnarray}
\displaystyle &&\int d\Omega O_{b}^{a}O_{d}^{+c}=\frac{1}{N_{c}}\delta
_{d}^{a}\delta _{b}^{c}, \\
\tau _{\mu }^{\pm }\tau _{\nu }^{\mp } &=&\delta _{\mu \nu }+i\eta _{\mu \nu
}^{a,\mp }\tau ^{a},\qquad \qquad \tau ^{a}\tau ^{b}=\delta
^{ab}+i\varepsilon ^{abc}\tau ^{c}.  \nonumber
\end{eqnarray}
Then, it is convenient to define the combinations of functions $D\left(
x^{2}\right) $ and $D_{1}\left( x^{2}\right) $ \cite{Mih93}
\begin{eqnarray}
A\left( x^{2}\right) &=&\delta _{\mu \rho }\delta _{\nu \sigma }\frac{D^{\mu
\nu ,\rho \sigma }(x)}{\left\langle 0\left| F_{\mu \nu }^{2}\right|
0\right\rangle ^{CI}}=D\left( x^{2}\right) +D_{1}\left( x^{2}\right) +\frac{1%
}{2}x^{2}\frac{\partial D_{1}\left( x^{2}\right) }{\partial x^{2}},
\label{AB_FF} \\
B\left( x^{2}\right) &=&4\frac{x_{\mu }x_{\rho }}{x^{2}}\delta _{\nu \sigma }%
\frac{D^{\mu \nu ,\rho \sigma }(x)}{\left\langle 0\left| F_{\mu \nu
}^{2}\right| 0\right\rangle ^{CI}}=D\left( x^{2}\right) +D_{1}\left(
x^{2}\right) +x^{2}\frac{\partial D_{1}\left( x^{2}\right) }{\partial x^{2}},
\nonumber
\end{eqnarray}
taking the boundary condition, $D(0)+D_{1}(0)=1$ and the asymptotic
conditions $D(\infty )=D_{1}(\infty )=0$. After direct, but cumbersome
calculations we come to the expressions for the functions $A$ and $B$:
\begin{eqnarray}
\displaystyle &&A(x^{2})=\frac{8}{\pi }N_{D}\int_{0}^{\infty }\
drr^{2}\int_{0}^{\infty }\ dt\left\{ \left[ \omega _{1}\left( z_{+}\right)
\omega _{1}\left( z_{-}\right) +\omega _{3}\left( z_{+}\right) \omega
_{3}\left( z_{-}\right) \right] \left( 3-4\sin ^{2}(\alpha _{z})\right)
-\right.  \label{AFF} \\
&&\left. -2\omega _{2}\left( z_{+}\right) \omega _{2}\left( z_{-}\right)
\left[ r^{2}x^{2}\left( 1-2\sin ^{2}(\alpha _{z})\right) -rx\left(
z_{+}\cdot z_{-}\right) \sin (2\alpha _{z})\right] \right\} ,\   \nonumber
\end{eqnarray}
\begin{eqnarray}
\displaystyle &&B(x^{2})=\frac{16}{\pi }N_{D}\int_{0}^{\infty }\
drr^{2}\int_{0}^{\infty }\ dt\left\{ \omega _{1}\left( z_{+}\right) \omega
_{1}\left( z_{-}\right) \left( 3-4\sin ^{2}(\alpha _{z})\right) -\right.
\label{BFF} \\
&&-\omega _{1}\left( z_{+}\right) \omega _{2}\left( z_{-}\right) \left[
z_{-}^{2}+2t_{-}^{2}\left( 1-2\sin ^{2}(\alpha _{z})\right) +2rt_{-}\sin
(2\alpha _{z})\right] -  \nonumber \\
&&-\omega _{2}\left( z_{+}\right) \omega _{1}\left( z_{-}\right) \left[
z_{+}^{2}+2t_{+}^{2}\left( 1-2\sin ^{2}(\alpha _{z}))\right) -2rt_{+}\sin
(2\alpha _{z})\right] +  \nonumber \\
&&\left. +\omega _{2}\left( z_{+}\right) \omega _{2}\left( z_{-}\right)
\left[ z_{+}^{2}z_{-}^{2}+2t_{+}t_{-}\left( z_{+}\cdot z_{-}\right) \left(
1-2\sin ^{2}(\alpha _{z})\right) +2rxt_{+}t_{-}\sin (2\alpha _{z})\right]
\right\} ,  \nonumber
\end{eqnarray}
where $\displaystyle z_{\pm }=\left( r,t_{\pm }\right) $, $\displaystyle %
t_{\pm }=t\pm \frac{x}{2}$, the forms $\ \omega _{1}\left( z\right) ,$ \ $%
\omega _{2}\left( z\right) ,$ and $\omega _{3}\left( z\right) $ are defined
in (\ref{OmFrms}), $N_{D}$ is the normalization factor
\begin{equation}
N_{D}^{-1}=6\int_{0}^{\infty }dyy^{3}\left( \omega _{1}^{2}\left( y\right)
+\omega _{3}^{2}\left( y\right) \right) ,  \label{FFnorm}
\end{equation}
and the phase factor
\[
\displaystyle\alpha _{z}=r\int_{-\frac{x}{2}}^{\frac{x}{2}}d\tau \varphi
_{g}\left( r^{2}+\left( t+\tau \right) ^{2}\right) ,
\]
reflects the presence of the $\hat{E}$ exponent in the definition of the
bilocal correlator. These expressions for the field strength correlators are
general for any field given in the form (\ref{CIanA}). The gauge invariance
of the functions $A(x^{2})$ and $B(x^{2})$ can explicitly be checked for
Exp.(\ref{AFF}, \ref{BFF}) by transforming, for example, the field $A_{\mu }$
from the singular to regular gauges, Eq. (\ref{Sin-RegTr}). The expressions
for $A(x^{2})$ and $B(x^{2})$ may be considered as generation functions to
obtain condensates of higher dimensions in the instanton model approach.
From a technical point of view this procedure is more convenient than their
direct calculations \cite{Mih93,DEM97}.

In the SI approximation the form factors $\ B^{I}(x^{2})=A^{I}(x^{2})$
reproduce the expression, Eq. (21) from \cite{DEM97}, for the
gauge-invariant correlator. As it has been shown in \cite{DEM97} (see also
\cite{Sim98}), in the SI approximation the term with the second Lorentz
structure, $D_{1}(x^{2})$, parameterizing the gluon correlator (\ref{Fld_Cor}
) does not appear. This fact is due to the specific topological structure
(self-duality) of the instanton solution. Both the Lorentz structures arise
in the r.h.s. of (\ref{Dgax}) if one takes into account the background fields.

The form factors $D(x^{2})$ and $D_{1}(x^{2})$ are determined numerically by
solving the equations (\ref{AB_FF}) and plotted in Fig. (\ref{Fig5}) in
coordinate space and in Fig. (\ref{Fig6}) in momentum space. The constant $%
\kappa $ defining the relative weight of $D$ functions $\left( D\left(
0\right) =\kappa ,D_{1}\left( 0\right) =1-\kappa \right) $ depends on the
background field strength parameter $\eta _{g}\rho $ and it is close to one
in the region of reasonable physical parameters. It is equal to $\kappa =1$
at $\left( \eta _{g}\rho \right) ^{2}=0$ (SI case), $\kappa =0.997$ at $%
\left( \eta _{g}\rho \right) ^{2}=0.1$ and $\kappa =0.926$ at $\left( \eta
_{g}\rho \right) ^{2}=3$. In all the cases the parameter $\kappa $ is close
to one in accordance to the fits of lattice data \cite{Megg98}. It means
that in the weak background field the dominant role of the $D(x^{2})$
function remains as in the SI case; it is close to the SI form at small
distances and exponentially rapidly decays at large distances. The $%
D_{1}(x^{2})$ function is small and positive everywhere, its behavior is
very sensitive not only to external field but also to the gauge phase factor
effect\footnote{%
In ref. \cite{Sim98}, in the calculations of the form factors $D,D_{1}$
based on the instanton-anti-instanton ansatz both the influence of the
physical vacuum on the instantons and the gluon correlators as an effect of
the $P-$ exponential factor on the form factors has been ignored. As a
result, the negative $D_{1}$ has been obtained in \cite{Sim98}. However,
both the facts are important in determining the correct norm and forms of
the form factors, in particular, in obtaining a small $D_{1}$}. In Appendix
A we show that the reason for smallness of the $D_{1}(x^{2})$ function is
``almost'' self-duality property of the CI solutions. Both the form factors
possess two zeros at large distances and at very large $x$ develop positive
asymptotics
\begin{equation}
D(x^{2}),D_{1}(x^{2})\sim |x|^{-3/4}(\eta _{g}\rho )^{4}\exp \left(
-0.473\left( \eta _{g}\left| x\right| \right) ^{3/2}\right) ,  \label{CorrAs}
\end{equation}
where $\eta _{g}\neq 0$ and the constant in the exponent is found
numerically.

In order to have contact with the QCD vacuum phenomenology and specify
further the instanton-induced model of the gluon correlator, let us discuss
the contributions of different terms in (\ref{CorrFull}) to the gluon
condensate

\begin{eqnarray}
&&\left\langle :F_{\mu \nu }[A^{CI}+b]\left( 0\right) F_{\mu \nu
}[A^{CI}+b]\left( 0\right) :\right\rangle =\left\langle F_{\mu \nu
}^{CI}\left( 0\right) F_{\mu \nu }^{CI}\left( 0\right) \right\rangle +
\label{GlCondFul} \\
&&+\left\langle :F_{b,\mu \nu }\left( 0\right) F_{b,\mu \nu }\left( 0\right)
:\right\rangle +\left\langle :\Delta F_{\mu \nu }[A^{CI}+b]\left( 0\right)
\Delta F_{\mu \nu }[A^{CI}+b]\left( 0\right) :\right\rangle ^{interf}.
\nonumber
\end{eqnarray}
The background contribution to the gluon condensate
\begin{equation}
\left\langle :F_{b,\mu \nu }\left( 0\right) F_{b,\mu \nu }\left( 0\right)
:\right\rangle =\left\langle F_{b}^{2}\right\rangle _{b}  \label{GlCondL}
\end{equation}
serves as a parameter of the model and, by assumption, is much smaller than
the CI contribution given by
\begin{equation}
\left\langle F_{\mu \nu }^{CI}\left( 0\right) F_{\mu \nu }^{CI}\left(
0\right) \right\rangle =32\pi ^{2}n_{c}N_{D}^{-1},  \label{GlCondCI}
\end{equation}
where $N_{D}^{-1}\approx 1$ (see Fig. \ref{Fig4}). The interference term
after averaging over relative color orientations and using relations (\ref
{ColAv}), (\ref{F-SchBB}) acquires the form
\begin{eqnarray}
&&\left\langle :F_{\mu \nu }[A^{CI}+b]F_{\mu \nu }[A^{CI}+b]:\right\rangle
^{interf}=  \label{GlConInt} \\
&=&\frac{N_{c}}{16\left( N_{c}^{2}-1\right) }\left( 32\pi ^{2}n_{c}\right)
\left\langle F_{b}^{2}\right\rangle _{b}\int_{0}^{\infty }dzz^{7}\varphi
_{g}^{2}\left( z^{2}\right) \Phi (z^{2}),  \nonumber
\end{eqnarray}
where $\Phi (z^{2})$ is defined in (\ref{Phi}) (the explicit forms of $\Phi
(z^{2})$ are outlined in Appendix B).

The interference term depends on two dimensionless parameters $\alpha
_{g}=\rho _{c}\eta _{g}$ and $\beta =\rho _{c}/R$ and the background field
condensate can be parameterized as $\displaystyle\left\langle
F_{b}^{2}\right\rangle _{b}\rho _{c}^{4}=\frac{9\left( N_{c}^{2}-1\right) }{%
a_{\Phi }N_{c}}\alpha _{g}^{3}\beta $. The instanton size $\rho _{c}$ comes
in the last formula with high power and leads to indefiniteness of the
factor of order 2 in the relation of the external field condensate to the
parameters $\alpha _{g}$ and $\beta $. To reduce this uncertainty, we can
use the physical information about the vacuum properties provided by the QCD
SRs and lattice QCD. Indeed, as it has been shown in \cite{DEM97}, in the SI
case there is a relation between the instanton size and the average
virtuality of quarks in the vacuum, Eq. (\ref{Q_Virt}). The value of the
average quark virtuality has been estimated in the QCD sum rule analysis, $%
\lambda _{q}^{2}=0.5\pm 0.05\quad {\rm GeV}^{2}$ in \cite{Piv91}; $\lambda
_{q}^{2}=0.4\pm 0.1\quad {\rm GeV}^{2}$ in \cite{BI82}, and from the lattice
QCD calculations $\lambda _{q}^{2}=0.55\pm 0.05\quad {\rm GeV}^{2}$ in \cite
{KSch87}. The relations (\ref{Q_Virt}) remain good approximation  in the CI
case if the external field does not strongly deform the instanton. Numerical
calculations of $\lambda _{g}^{2}$, defined in (\ref{Q_Virt}), lead to
estimates
\begin{equation}
\lambda _{g}^{2}=4.8\frac{1}{\rho _{c}^{2}}~(\alpha _{g}^{2}=0),~~%
\displaystyle\lambda _{g}^{2}=5.7\frac{1}{\rho _{c}^{2}}~(\alpha _{g}^{2}=1).
\label{Q_Virt3}
\end{equation}
We show below that physically motivated background field has the strength
parameter $\alpha _{g}<1$ and thus the value of $\lambda _{g}^{2}$ increases
not more than $20\%$.

The relation for $\lambda _{q}^{2}$ in (\ref{Q_Virt}) can be used to get the
scale for the background field condensate
\[
\displaystyle\left\langle F_{b}^{2}\right\rangle _{b}=\frac{9\left(
N_{c}^{2}-1\right) }{4N_{c}a_{\Phi }}\left( \lambda _{q}^{2}\right)
^{2}\alpha _{g}^{3}\beta
\]
and we accept in the following
\[
\lambda _{q}^{2}=0.5\quad {\rm GeV}^{2}.
\]
What is the expected range for the parameters $\alpha _{g}$ and $\beta $? By
analogy with the instanton liquid vacuum, the parameter $\beta $ can be
interpreted as a ratio of the instanton size to the inter-instanton distance
and it is adjusted as $\beta \approx 1/3$. Then, the estimate of the upper
limit for the strength parameter $\alpha _{g}$ follows from the assumption
that the contribution of the background field to the total gluon condensate $%
\left\langle 0\left| F^{2}\right| 0\right\rangle _{total}\approx 1$ {\rm GeV}%
$^{4}$ \cite{Naris95} is quite small. This assumption reduces the influence
of the model dependent part of correlator and leads to the bound $\alpha
_{g}<1$, or for the dimensional parameter, $\eta _{g}^{2}=\alpha
_{g}^{2}\lambda _{q}^{2}/4$, $\eta _{g}<0.35$ {\rm GeV}. In Fig. (\ref{Fig8}%
) we present the values of the interference term as a two parametric plot
and see that its contribution to the gluon condensate is small if the
short-range and large-scale fluctuations are well separated: $\alpha _{g}<1$
and $\beta \ll 1$. For completeness in Appendix B we present the small
interference contributions to the functions $A(x^{2})$ and $B(x^{2})$.

Thus, we construct the model of the gluon correlators. Within this model the
functions $A(x^{2})$ and $B(x^{2})$ are the sum of the short-range instanton
induced contributions (\ref{AFF}) and (\ref{BFF}) multiplied by the weight
factor $32\pi ^{2}n_{c}/\left\langle 0\left| F^{2}\right| 0\right\rangle
_{total}$ and the long-range contribution (\ref{CorrLB}) modeled by Exp. (%
\ref{B}) with the weight factor $\left\langle F_{b}^{2}\right\rangle
_{b}/\left\langle 0\left| F^{2}\right| 0\right\rangle _{total}$. The
parameters of the model are the average instanton size $\rho _{c}\approx 0.3$
{\rm fm}, the effective instanton density $n_{c}\approx 1$ {\rm fm} $^{-4}$,
the strength $\left\langle F_{b}^{2}\right\rangle _{b}\leq 32\pi ^{2}n_{c}$
and the correlation length $R\approx 3\rho _{c}$ of the background field.
The first two parameters are estimated within the instanton liquid models,
being in the dilute liquid limit expressed through the vacuum averages: $%
\rho _{c}^{2}=2\lambda _{q}^{-2}$ and $n_{c}=\left( 2\pi I/N_{c}\right)
\left( \left| \left\langle \overline{q}q\right\rangle \right| ^{2}/\lambda
_{q}^{2}\right)$, where the numerical constant $I\approx 0.6$ \cite{DoLaur98}%
. The latter relation is a consequence of the gap equation \cite{DP84}. The
form of the short-range correlator is defined by $\rho _{c}$ at small
distances and by the long-scale parameter $\eta _{g}$ at large distances.
The form of the long-range correlator at large distances can be motivated by
the results obtained in the dual effective model of QCD \cite{BBDV}, where
they have exponential decrease (modulo powers) similar to exponential ansatz
in (\ref{B}). Basing on the results of the dual model and lattice
measurements one can expect that $A(x^{2})\approx B(x^{2})$ for the
long-range part of the correlator.

The field-strength correlators have been studied on the lattice in \cite
{DiGi97}, \cite{Bali97}. There, the following two combinations of form
factors have been measured:
\begin{eqnarray*}
D_{\bot }\left( x^{2}\right) &=&D\left( x^{2}\right) +D_{1}\left(
x^{2}\right) , \\
D_{||}\left( x^{2}\right) &=&D\left( x^{2}\right) +D_{1}\left( x^{2}\right)
+x^{2}\frac{\partial D_{1}\left( x^{2}\right) }{\partial x^{2}},
\end{eqnarray*}
where $D_{\bot }\left( x^{2}\right) =2A\left( x^{2}\right) -B\left(
x^{2}\right) $ and $D_{||}\left( x^{2}\right) =B\left( x^{2}\right) $ in
terms of the combinations defined in (\ref{AB_FF}). The lattice measurements
of the field strength correlators are also obtained with the straight line
path in the Schwinger exponent. The direct comparison of the model
calculations with the lattice data is a delicate problem, since the used
parameterization is rather conventional in separating the residual
perturbative tail (divergent term $\sim x^{-4}$ ) from the non-perturbative
part (pure exponential finite term). As a result, the fits with {\it ad hoc }%
chosen parameterizations are very unstable with respect to the extraction of
the quantities of physical interest: the correlation lengths, the gluon
condensate, {\it etc.} \cite{Megg98}. The reason is that the perturbative
part is strongly divergent (as it is seen from lattice data), its
contribution at small distances would be strongly dependent on the
parameterization procedure. On the other hand, by construction we calculate
the non-perturbative part of the correlators with perturbative contributions
subtracted. In future, it would be quite desirable to make a new fit to the
lattice data using the Eqs. (\ref{AFF}), (\ref{BFF}), as an input for a
non-perturbative part of the correlators.

At the present stage, we restrict ourselves only to a few qualitative
remarks. In ref. \cite{Megg98}, the range of values of some physical
quantities was discussed which can be fitted from the lattice data according
to different parameterizations. Namely, the correlation length, the gluon
condensate and the normalization of form factors have been analyzed. As it
has been noted above the normalization of form factors $\kappa $, consistent
with small $D_{1}\left( x^{2}\right) $, is in agreement with the instanton
model and that the value of the gluon condensate serves as a free model
parameter. The values of the gluon condensate extracted from the fits to
lattice data are very sensitive to the parameterization used, being within
the interval $<:\frac{\alpha _{s}}{\pi }~F_{\mu \nu }^{a}F_{\mu \nu
}^{a}:>=\left( 0.005-0.03\right) $ {\rm GeV}$^{4}$. In the lattice
``full-QCD'' fit of an average correlation length $l_{G}$ of the gluon
strength, defined as \cite{Megg98}
\begin{equation}
l_{G}=\frac{1}{D\left( 0\right) }\int_{0}^{\infty }dxD\left( x^{2}\right) ,
\label{lG}
\end{equation}
is in the range $l_{G}\approx 0.35-0.45$ {\rm fm }with lattice quark mass $%
am_{q}=0.01$ and $l_{G}\approx 0.3-0.4$ {\rm fm }with $am_{q}=0.02$, where $a
$ is the lattice unit. One can expect, following linear extrapolation, that
in the chiral limit $am_{q}\rightarrow 0$,~$l_{G}\approx 0.4-0.5$ {\rm fm}.
Now let us omit an important but unsolved problem about the difference of
lattice and CI renormalization schemes and norms, to draw rather a rough
comparison of the corresponding results. The instanton model predicts for
the same quantity: $l_{G}=0.43$ {\rm fm }at $\rho _{c}\eta _{g}=0$ (SI), $%
l_{G}=0.37$ {\rm fm }at $\left( \rho _{c}\eta _{g}\right) ^{2}=1$, $%
l_{G}=0.31$ {\rm fm }at $\left( \rho _{c}\eta _{g}\right) ^{2}=3$. Thus, the
predictions of the instanton model, under the considered condition $\alpha
_{g}\equiv \rho _{c}\eta _{g}<1$, are in qualitative agreement with
information extracted from the lattice data.

\section{Conclusions}

The instanton model provides a way of constructing the nonlocal vacuum
condensates. We have obtained the expressions for the nonlocal gluon $%
<:TrF^{\mu \nu }(x)\hat{E}(x,y)F^{\rho \sigma }(y)\hat{E}(y,x):>$ correlator
beyond the single instanton (SI) approximation\cite{DEM97}. They have
consistent properties at short as well as at large distances. The model
constructed predict the behaviour of nonperturbative part of gluon
correlation functions in the short and intermediate region assuming that it
is dominated by instanton vacuum component.

To this goal, we have suggested that the instanton $A_{\mu }^{CI}(x)$ is
developed in the physical vacuum field $b_{\mu }(x)$ interpolating
large-scale vacuum fluctuations. We have found that at small distances the
instanton field dominates, and at large distances it decreases
exponentially. We did not assume any particular properties of the long-wave
vacuum field $b_{\mu }(x)$ but managed to reduce the effect to certain
phenomenological quantities, namely, the correlation function $\widetilde{B}
(x^{2})$ determined by its strength $\left\langle F_{b}^{2}\right\rangle
_{b} $ and the correlation length $R$. Within this model, by averaging over
random color vector orientations of the background field with respect to the
fixed instanton field orientation, we have found equation (\ref{E-LeqAv1})
governing the deformation of the instanton under the influence of the weak
background vacuum field. Following Affleck idea we have shown that, to
stabilize the instanton, we need to put constraints on the system. Next, we
have found the constraint independent asymptotics of the instanton solution
at large distances, given by Eqs. (\ref{CIanz}) and (\ref{AsSol}), where it
is exponentially suppressed $\displaystyle A_{\mu }^{CIa}(x)\sim 2\overline{
\eta }_{\nu \mu }^{a}x_{\nu }\frac{\left( \rho_c \eta _{g}\right) ^{2}}{%
x^{2}\left| x\right| ^{3/4}}\exp \left[ -\frac{2}{3}\left( \eta _{g}\left|
x\right| \right) ^{3/2}\right] $ unlike the powerful decreasing SI. It is
important to note that the form of this asymptotics is also independent on
the model for the background field and the driven parameter $\displaystyle
\eta _{g}\sim \left( \frac{N_{c}}{9\left( N_{c}^{2}-1\right) }R\left\langle
F_{b}^{2}\right\rangle _{b}\right) ^{\frac{1}{3}}$ only weakly depends on
it. Assuming that the external field is weak, the CI profile function is
close to SI profile at distances smaller than $\rho _{c}$ and it decreases
exponentially at distances larger than $\eta _{g}^{-1}$ (see (\ref{Fig1})).
In particular, this result means that the leading interaction term of a
widely separated instanton-anti-instanton pair in physical vacuum decays
exponentially with separation and differs from dipole interaction term found
previously in unconstrained model. The knowledge of the
constraint-independent parts of CI allowed us to construct the solution in
the ansatz form (\ref{CIanA}) with the profile functions (\ref{CIprofM}) - (%
\ref{CIprofSW}). As it is seen from Figs. (\ref{Fig2}) - (\ref{Fig4}) the
profile of the CI and its action are practically independent of the choice
of the ansatz if the interference parameter is in the region, $\rho_{c}\eta
_{g}<1$, where our considerations are justified.

Then, for an arbitrary classical gauge field of the form $A_{\mu
}^{CIa}(x)=2 \overline{\eta }_{\nu \mu }^{a}x_{\nu }\varphi _{g}\left(
x^{2}\right) $,\ we have found the expressions (\ref{AFF}) and (\ref{BFF})
for the combinations of form factors $D\left( x^{2}\right) $ and $%
D_{1}\left( x^{2}\right) $, which parameterize the gauge-invariant gluon
field strength correlator. These expressions generalize the previously known
expressions for the SI model \cite{DEM97}. The correlators have been
calculated numerically. As it turns out, at a reasonable set of parameters,
guaranteeing the smallness of the large-scale vacuum field fluctuations, the
$D\left( x^{2}\right) $ structure is close to the SI induced function with
the exponential asymptotics being developed at large distances\footnote{%
The similar behavior is expected for the quark correlator in the physical
vacuum\cite{DEMM99}.}. At the same time, the $D_{1}\left( x^{2}\right) $
structure is about two orders smaller than the $D\left( x^{2}\right) $
function at any reasonable choice of the parameter $\rho _{c}\eta _{g}$. As
it is explained in the Appendix A, the reason is that in the dilute vacuum
the CIs are ``almost'' self-dual. The relative strength of the form factors $%
D\left( x^{2}\right) $ and $D_{1}\left( x^{2}\right) $\ is very sensitive to
the accepted physical picture. The lattice data are in qualitative agreement
with predictions of the constrained instanton model. It means, in
particular, that in the interpretation of the lattice data more justified
parameterization for the correlation functions can be used. It allows one to
extract from data the values of physical interest and separate the
perturbative tail from the nonperturbative contribution. Moreover, due to
fast decay of the CI induced part of correlators, the exponential decay
observed in lattice calculations can be attributed to the background
component of the vacuum field, or be described by some other field
theoretical approaches \cite{BBDV,DEJ}. From the other side, the SI model is
inconsistent with large distance behavior. The nonperturbative part of the
functions $A(x^{2})$ and $B(x^{2})$ are the sum of short-range instanton
induced contributions (\ref{AFF}) and (\ref{BFF}), multiplied by the weight
factor $n_{c}32\pi ^{2}/\left\langle 0\left| F^{2}\right| 0\right\rangle
_{total}$, and the long-range contribution (\ref{CorrLB}), modeled by
exponentially decreasing function (\ref{B}) with the weight factor $%
\left\langle F_{b}^{2}\right\rangle _{b}/\left\langle 0\left| F^{2}\right|
0\right\rangle _{total}$.

The constrained instanton model introduces two characteristic scales
(correlation lengths). One is related to short distance behavior of the
correlation functions and another with long range distance behavior. The
first one, $\lambda _{g}^{-1},$ is predictable and expressed in terms of
physical quantities. In SI approximation, given by Eqs. (\ref{Q_Virt}) and (%
\ref{Q_Virt2}), it is proportional to the instanton size, and gains small
negative corrections due to the background, Eq. (\ref{Q_Virt3}). As to large
scale it is out of the scope of our model, and we can only physically relate
it to the confinement size or extract it from long distance asymptotics of
lattice calculations. The microscopic description of the long distance
background field needs other considerations not examined in the present work.

The calculations have been performed in a gauge - invariant manner by using
the expressions for the instanton field in the axial gauge. The behavior of
the correlation functions demonstrates that in the single constrained
instanton approximation the model of nonlocal condensates can well reproduce
the behavior of the functions at short and intermediate distances, while the
large-scale asymptotics is dominated by the background field. It would be
quite desirable to make a fit to the lattice data using as an input the
instanton induced correlators. The important question concerning the
interacting ensemble of the constrained instantons has also to be postponed
for another specific work.

\centerline{\bf Acknowledgments} \vspace{2mm} The authors are grateful to
N.I. Kochelev, L.Tomio for fruitful discussions of the results. We thank Dr.
E. Meggiolaro for conversation on the lattice QCD data on gluon correlators.
One of us (A.E.D.) thanks the colleagues from Instituto de F\'{i}sica
Te\'{o}rica, UNESP, (S\~{a}o Paulo) for their hospitality and interest in
the work. This investigation has been supported in part by the Russian
Foundation for Fundamental Research (RFFR) 96-02-18096 (A.E.D., S.V.E),
96-02-18097 (A.E.D.) and 96-02-17631 (S.V.M.), St. - Petersburg center for
fundamental research grant: 97-0-6.2-28 (A.E.D.) and INTAS 93-283-ext
(S.V.E.) . A.E.D. also thanks partial support received from Funda\c{c}\~{a}o
de Amparo \`{a} Pesquisa do Estado de S\~{a}o Paulo (FAPESP) and from
Conselho Nacional de Desenvolvimento Cient\'{i}fico e Tecnol\'{o}gico do
Brasil.

\begin{appendix}
\appendix
\section{Appendix}

The question may arise why the short-range vacuum contributions to the
$D_{1}\left( x^{2}\right) $ structure are negligible. As it was
found in \cite {DEM97} in the SI case $D_{1}\left( x^{2}\right)
=0$ due to self-duality of instanton solution. The CIs are not
self-dual and contribute to $D_{1}\left( x^{2}\right) $, but to
what extent the self-duality is violated? We are going to show
that for the reasonable set of parameters and ansatze assumed the
CIs are ``almost'' self-dual and this is the reason for smallness of
$D_{1}\left( x^{2}\right) $. On the contrary if lattice simulations
detected very big contribution to $D_{1}\left( x^{2}\right) $, it
would mean that self-duality is lost and there is no chances to
save instantons as individual objects in the QCD vacuum. From the
point of view of our model, any essential contributions to
$D_{1}\left( x^{2}\right) $ can arise only from the large-scale
vacuum fluctuations.

The dual field strength $\displaystyle\widetilde{F}_{\mu \nu }^{CI,a}\left(
x\right) =\frac{1}{2}\epsilon _{\mu \nu \rho \sigma }F_{\mu \nu
}^{CI,a}\left( x\right) $, where $F_{\mu \nu }^{CI,a}\left( x\right) $ is
defined
in (\ref{CIanF}), can be expressed in the form
\begin{equation}
\widetilde{F}_{\mu \nu }^{CI,a}\left( x\right) =4\left[ \overline{\eta }
_{\mu \nu }^{a}\widetilde{\omega }_{1}\left( x\right) +\left( x_{\mu }
\overline{\eta }_{\nu \rho }^{a}-x_{\nu }\overline{\eta }_{\mu \rho
}^{a}\right) x_{\rho }\widetilde{\omega }_{2}\left( x\right) \right] ,
\label{F_Self_D}
\end{equation}
with forms $\widetilde{\omega }\left( x\right) $
\begin{equation}
\widetilde{\omega }_{1}\left( x\right) =\varphi _{g}\left( x^{2}\right)
+x^{2}\frac{\partial \varphi _{g}\left( x^{2}\right) }{\partial x^{2}}
,\qquad \widetilde{\omega }_{2}\left( x\right) =\omega _{2}\left( x\right)
=\varphi _{g}^{2}\left( x^{2}\right) +\frac{\partial \varphi _{g}\left(
x^{2}\right) }{\partial x^{2}}.
\end{equation}
Let us consider the difference of the field strengths given in the regular
at zero gauge
\begin{equation}
F_{\mu \nu }^{CIa}\left( x\right) -\widetilde{F}_{\mu \nu }^{CIa}\left(
x\right) =4\left[ \overline{\eta }_{\mu \nu }^{a}+2\frac{\left( x_{\mu }
\overline{\eta }_{\nu \rho }^{a}-x_{\nu }\overline{\eta }_{\mu \rho
}^{a}\right) x_{\rho }}{x^{2}}\right] \omega _{2}^{reg}\left( x\right) x^{2}.
\label{Self_D_Cond}
\end{equation}
The self-duality condition $F_{\mu \nu }^{CIa}\left( x\right) -\widetilde{F}
_{\mu \nu }^{CIa}\left( x\right) =0$ is satisfied for the SI case,
where $\ x^{2}\omega _{2}^{reg,I}\left( x\right) =0$ (see Eq.
(\ref{OmFrmsReg})). Comparing Eqs. (\ref{CIanF}) and
(\ref{F_Self_D}) with Eq.(\ref{Self_D_Cond}) we can consider the
condition $$\left| \omega _{2}^{reg,CI}\left( x\right) \right|
x^{2}<<\left| \omega _{1}^{reg,CI}\left( x\right) \right| \sim
\left| \omega _{1}^{reg,I}\left( x\right) \right| $$
 as a criterion
indicating that the field is ``almost'' self-dual. It can be
checked numerically, that it is really the case at reasonable
choice of background field strength parameter $\alpha _{g} < 1$
and all forms of ansatze for CI. At the same time at larger values
of parameter $\alpha _{g}\simeq 4\div 6 $ the inequality is
not fulfilled. Thus we show that assuming diluteness of the vacuum the
CIs are ``almost'' self-dual solutions and, as a consequence,
contribute very small to the $D_{1}\left( x^{2}\right) $ structure
of the vacuum gluon field strength correlators.

\section{Appendix}
 The function $\Phi (z^{2})$ for the different forms of the
form factor $\widetilde{B}\left( x^{2}\right) $:
\begin{equation}
\Phi _{G}(z^{2})=\frac{2}{3y^{4}}\left[ 2\sqrt{\pi }y^{3}
erf\left( y\right) -3y^{2}+1-\left( 1-2y^{2}\right) \exp \left(
-y^{2}\right) \right] ,\qquad \mbox{Gaussian,} \label{PhiG}
\end{equation}
\begin{equation}
\Phi _{M}(z^{2})=\frac{2}{3y^{4}}\left[ 4y^{3}\arctan \left(
y\right) +y^{2}-\left( 1+3y^{2}\right) \ln \left( 1+y^{2}\right)
\right] ,\qquad \mbox{monopole,}  \label{PhiM}
\end{equation}
and
\begin{equation}
\Phi _{E}(z^{2})=\frac{4}{3y^{4}}\left[ 2y^{3}-3y^{2}+6-6\left(
1+y\right) \exp \left( -y\right) \right] ,\qquad
\mbox{exponential,}  \label{PhiE}
\end{equation}
where $y=z/R$, $R$ being the correlation length of the large-scale
vacuum field.

 For completeness we present here the small
interference contribution to the functions $A(x^{2})$ and
$B(x^{2})$
\begin{eqnarray}
\displaystyle &&A^{interf}(x^{2})=N_{D}\left\langle 0\left| F_{b}^{2}\right|
0\right\rangle \frac{N_{c}}{6\left( N_{c}^{2}-1\right) }\int_{0}^{\infty }\
drr^{2}\int_{0}^{\infty }\ dt\Phi (z_{+},z_{-})\varphi _{g}\left(
z_{+}^{2}\right) \varphi _{g}\left( z_{-}^{2}\right) \cdot   \label{Ainterf}
\\
&&\cdot \left( z_{+}\cdot z_{-}\right) \left\{ \left( z_{+}\cdot
z_{-}\right) \left( 3-4\sin ^{2}(\alpha _{z})\right) -xr\sin (2\alpha
_{z})\right\} ,  \nonumber
\end{eqnarray}
\begin{eqnarray*}
\displaystyle &&B^{interf}(x^{2})=N_{D}\left\langle 0\left| F_{b}^{2}\right|
0\right\rangle \frac{N_{c}}{6\left( N_{c}^{2}-1\right) }\int_{0}^{\infty }\
drr^{2}\int_{0}^{\infty }\ dt\Phi (z_{+},z_{-})\varphi _{g}\left(
z_{+}^{2}\right) \varphi _{g}\left( z_{-}^{2}\right) \cdot  \\
&&\cdot r^{2}\left\{ 2\left( z_{+}\cdot z_{-}\right) \left( 2-3\sin
^{2}(\alpha _{z})\right) -xr\sin (2\alpha _{z})\right\} ,
\end{eqnarray*}
where
\[
\Phi (z_{+},z_{-})=4\int_{0}^{1}d\alpha \int_{0}^{1}d\beta \alpha \beta
\widetilde{B}\left[ \left( \alpha z_{+}-\beta z_{-}\right) ^{2}\right] ,
\]
$z_{\pm }=\left( r,t\pm \frac{x}{2}\right) $, $z=\left( r,t\right) ,$ and
$\varphi _{g}\left( z^{2}\right) $ is defined in (\ref{CIanA}). In deriving
the above expressions we used the Schwinger - Fock gauge for background
field, Eqs. (\ref{F-Sch}) - (\ref{CorrLB}), and neglected the derivative
$\widetilde{D}_{1}^{\prime }(x^{2})$ in comparison with the
function $\widetilde{B}(x^{2})$ itself. In fact, it is
consistent with $\widetilde{D}_{1}\left( x^{2}\right) =0$ that
follows from the lattice data \cite{DiGi97} and instanton
calculations \cite{DEM97}. We find that the interference
contributions to the correlators are very small in absolute
value, have shorter correlation length comparing with the CI contributions (
\ref{AFF}), (\ref{BFF}) and do not lead to valuable appearance of the $
D_{1}\left( x^{2}\right) $ structure.\

\end{appendix}

\newpage FIGURE CAPTIONS:

\bigskip

Fig. 1: The constrained instanton profile functions $x^{2}\varphi_{g}(\left|
x\right| /\rho )$ (\ref{CIanA}), corresponding to the ansatz (\ref{CIprofSW}%
), at different values of the parameter $(\rho\eta _{g})^{2}$: $(\rho\eta
_{g})^{2}=0$ solid line (instanton case), $(\rho\eta _{g})^{2}=0.5$- short
dashed line, $(\rho\eta_{g})^{2}=3$ - long dashed line.

Fig. 2: The instanton (solid line) and constrained instanton profile
functions $x^{2}\varphi _{g}(\left| x\right| /\rho )$ (\ref{CIanA})
corresponding to different ansatze: (\ref{CIprofM}) dotted line, (\ref
{CIprofW}) - short dashed line, (\ref{CIprofSW}) - long dashed line, at the
large value of the parameter $(\rho \eta _{g})^{2}=3$.

Fig. 3: The density action of the instanton (solid line) and constrained
instanton $G^2(x^2)\equiv \rho^4/192 F_{\mu \nu }^{Ia}\left( x\right) F_{\mu
\nu }^{Ia}\left( x\right) $, Eq. (\ref{ActDens}), as a function of $\left|
x\right| /\rho$ corresponding to different ansatze: (\ref{CIprofM}) dotted
line, (\ref{CIprofW}) - short dashed line, (\ref{CIprofSW}) - long dashed
line, at the value of the parameter $(\rho\eta _{g})^{2}=1$.

Fig. 4: The classical action of the instanton (solid line) and the
constrained instanton$S_{CI}$, Eq. (\ref{SCI}), as a function of $\rho\eta
_{g}$ corresponding to different ansatze: (\ref{CIprofM}) dotted line, (\ref
{CIprofW}) - short dashed line, (\ref{CIprofSW}) - long dashed line. The
action is given in units of $8\pi ^{2}/g^{2}$.

Fig. 5: The form factors $D$ (top lines) and $D_{1}$ (bottom lines) (all
normalized by $D(0)$) versus physical distance $x$, for the instanton size $%
\rho=0.3$ fm and parameters $(\rho\eta _{g})^{2}=0$ (solid lines) and $%
(\rho\eta _{g})^{2}=1 $ (dashed lines).

Fig. 6: The form factor $\tilde D(p)$ as a function of $\rho |p|$
corresponding to the ansatz (\ref{CIprofSW}), at different values of the
parameter $(\rho\eta _{g})^{2}$ : $(\rho\eta _{g})^{2}=0$ solid line
(instanton case), $(\rho\eta _{g})^{2}=0.5$\ - short dashed line, $%
(\rho\eta_{g})^{2}=3$ - long dashed line.

Fig. 7: The interference term contribution to the gluon condensate
normalized by the instanton contribution $F_{\mu \nu }[A^{I}]F_{\rho \sigma
}[A^{I}]=n_{c} 32\pi ^{2}$ as function of the large-scale vacuum fluctuation
correlation length $1/\beta = R/\rho_c$ and its strength parameter $%
\alpha_g=\eta _{g}\rho_c.$

\bigskip 
\qquad \qquad\

\begin{figure}[tbp]
\vspace*{0.5cm} \epsfxsize=15cm \epsfysize=15cm \centerline{\epsfbox{fig.1}}
\caption[dummy0]{ The constrained instanton profile functions $x^{2} \protect%
\varphi _{g}(\left| x\right| /\protect\rho )$ (\ref{CIanA}), corresponding
to the ansatz (\ref{CIprofSW}), at different values of the parameter $(%
\protect\rho\protect\eta _{g})^{2}$ : $(\protect\rho\protect\eta _{g})^{2}=0$
solid line (instanton case), $(\protect\rho\protect\eta _{g})^{2}=0.5$\ -
short dashed line, $(\protect\rho\protect\eta_{g})^{2}=3$ - long dashed
line. }
\label{Fig1}
\end{figure}

\begin{figure}[tbp]
\vspace*{0.5cm} \epsfxsize=15cm \epsfysize=15cm \centerline{\epsfbox{fig.2}}
\caption[dummy0]{ The instanton (solid line) and constrained instanton
profile functions $x^{2}\protect\varphi _{g}(\left| x\right| /\protect\rho )$
(\ref{CIanA}) corresponding to different ansatze: (\ref{CIprofM}) dotted
line, (\ref{CIprofW}) - short dashed line, (\ref{CIprofSW}) - long dashed
line, at the large value of the parameter $(\protect\rho \protect\eta
_{g})^{2}=3$. }
\label{Fig2}
\end{figure}

\begin{figure}[tbp]
\vspace*{0.5cm} \epsfxsize=15cm \epsfysize=15cm \centerline{\epsfbox{fig.3}}
\caption[dummy0]{ The density action of the instanton (solid line) and
constrained instanton $G^2(x^2)\equiv \protect\rho^4/192 F_{\protect\mu
\protect\nu }^{Ia}\left( x\right) F_{\protect\mu \protect\nu }^{Ia}\left(
x\right) $, Eq. (\ref{ActDens}), as a function of $\left| x\right| /\protect%
\rho$ corresponding to different ansatze: (\ref{CIprofM}) dotted line, (\ref
{CIprofW}) - short dashed line, (\ref{CIprofSW}) - long dashed line, at the
value of the parameter $(\protect\rho\protect\eta _{g})^{2}=1$. }
\label{Fig3}
\end{figure}

\begin{figure}[tbp]
\vspace*{0.5cm} \epsfxsize=15cm \epsfysize=15cm \centerline{\epsfbox{fig.4}}
\caption[dummy0]{ The classical action of the instanton (solid line) and
constrained instanton$S_{CI}$, Eq. (\ref{SCI}), as a function of $\protect%
\rho\protect\eta _{g}$ corresponding to different ansatze: (\ref{CIprofM})
dotted line, (\ref{CIprofW}) - short dashed line, (\ref{CIprofSW}) - long
dashed line. The action is given in units of $8\protect\pi ^{2}/g^{2}$. }
\label{Fig4}
\end{figure}

\begin{figure}[tbp]
\vspace*{0.5cm} \epsfxsize=15cm \epsfysize=15cm \centerline{\epsfbox{fig.5}}
\caption[dummy0]{ The form factors $D$ (top lines) and $D_{1}$ (bottom
lines) (all normalized by $D(0)$) versus physical distance $x$, for the
instanton size $\protect\rho=0.3$ fm and parameters $(\protect\rho \protect%
\eta _{g})^{2}=0$ (solid lines) and $(\protect\rho\protect\eta _{g})^{2}=1$
(dashed lines). }
\label{Fig5}
\end{figure}

\begin{figure}[tbp]
\vspace*{0.5cm} \epsfxsize=15cm \epsfysize=15cm \centerline{\epsfbox{fig.6}}
\caption[dummy0]{The form factor $\tilde D(p)$ as a function of $\protect%
\rho |p|$ corresponding to the ansatz (\ref{CIprofSW}), at different values
of the parameter $(\protect\rho\protect\eta _{g})^{2}$ : $(\protect\rho%
\protect\eta _{g})^{2}=0$ solid line (instanton case), $(\protect\rho\protect%
\eta _{g})^{2}=0.5$\ - short dashed line, $(\protect\rho\protect\eta%
_{g})^{2}=3$ - long dashed line.}
\label{Fig6}
\end{figure}

\begin{figure}[tbp]
{\protect\vspace*{0.5cm} \epsfxsize=15cm \epsfysize=15cm
\centerline{\epsfbox{fig.7}}}
\caption[dummy0]{ The interference term contribution to the gluon condensate
normalized by the instanton contribution $F_{\protect\mu \protect\nu
}[A^{I}]F_{\protect\rho \protect\sigma }[A^{I}]=n_{c}32\protect\pi ^{2}$ as
function of the large-scale vacuum fluctuation correlation length $1/\protect%
\beta =R/\protect\rho _{c}$ and its strength parameter $\protect\alpha _{g}=%
\protect\eta _{g}\protect\rho _{c}.$ }
\label{Fig8}
\end{figure}

\end{document}